\documentclass[onecolumn,a4paper]{article}

\usepackage[english]{babel}
\usepackage{amsthm,amsmath,amssymb,amsfonts,amsopn}
\usepackage{color}
\usepackage{graphicx,subfigure,graphics,epstopdf,float}
\graphicspath{{figures/}}
\usepackage{enumerate}
\usepackage{algorithmic}
\usepackage{makecell}
\usepackage{booktabs}
\usepackage{hyperref}
\usepackage{soul}
\setstcolor{red}
\ifpdf
  \DeclareGraphicsExtensions{.eps,.pdf,.png,.jpg}
\else
  \DeclareGraphicsExtensions{.eps}
\fi

\newtheorem{remark}{Remark}

\newtheorem{thm}{Theorem}
\newtheorem{assum}{Assumption}
\newtheorem{defn}{Definition}
\newtheorem{lem}{Lemma}
\newtheorem{cor}{Corrollary}

\DeclareMathOperator{\q}{\mathbf{q}}

\DeclareMathOperator*{\E}{\mathrm{\bf E}}
\newcommand{\sys}[4]{\left[
\begin{array}{c|c}
#1 & #2 \\ \hline #3 & #4
\end{array}
\right]}

\newcommand{\hlt}[1]{{#1}}
\newcommand{\hltb}[1]{#1}

\newcommand{\msis}{mean-square {input-output} stability }

\begin{document}
\title{Mean-Square \hlt{Input-Output} Stability and Stabilizability of a Networked Control System with Random Channel Induced Delays}
\author{Weizhou Su\thanks{School of Automation Science and Engineering, South China University of Technology,
Guangzhou, China. \href{mailto:wzhsu@scut.edu.cn}{wzhsu@scut.edu.cn},
\href{mailto:l.junhui@mail.scut.edu.cn}{l.junhui@mail.scut.edu.cn},
\href{mailto:aujylu@scut.edu.cn}{aujylu@scut.edu.cn}}
\and Junhui Li\footnotemark[1]
\and Jieying Lu\footnotemark[1]}
\date{}



\maketitle

\begin{abstract}
This work mainly investigates the \msis and stabilizability for a single-input single-output (SISO) networked linear feedback system. The control signal in the networked system is transmitted over an unreliable channel. In this unreliable channel, the data transmission times, referred to as channel induced delays, are random values and the transmitted data could also be dropout with certain probability. The channel induced delays and packet dropout are modeled by an independent and identically distributed (i.i.d.) stochastic process with a fixed probability mass function (PMF). It is assumed that the transmitted data are with time stamps. At the channel terminal, a linear combination of data received at one sampling time is applied to the plant of the networked feedback system as a new control signal. To describe the uncertainty in the channel, a concept so called {\em {frequency response of variation}} is introduced for the unreliable channel. \hlt{With the given linear receiving strategy,} a \msis criterion is established in terms of the {{frequency response of variation}} of the unreliable channel for the networked feedback system. { {It is shown by this criterion that the \msis is determined by the interaction between the {{frequency response of variation}} and the nominal feedback system. In the \msis of the system, the role played by the random channel induced delays is \hltb{the} same as that played by a colored additive noise in an additive noise channel with a signal-to-noise ratio constraint}}. Moreover, the mean-square input-output stabilizability via output feedback is studied for the networked system. When the plant in the networked feedback system is minimum phase, an analytic necessary and sufficient condition is presented for its mean-square
input-output stabilizability. It \hltb{turns} out that the stabilizability is only determined by the interaction between the {{frequency response of variation}} of the channel and unstable poles of the plant. Finally, numerical examples are given to illustrate our results.
\end{abstract}



\section{Introduction}\label{Sec:Introduction}

Networked feedback control systems are known as spatially distributed control systems in which signals are exchanged over communication network \cite{Antsaklis2007}. In the past few decades, a huge amount of
research interests has been attracted on the analysis and design of these systems due to great advantages
in networked control systems, such as ease for installation and maintenance, reduced system wiring, and
resources sharing etc. Typical application examples of networked control systems can be found in a broad
range of areas such as: automotive industry \cite{Sandberg2015}, autonomous underwater vehicles and
unmanned air vehicles \cite{Jawhar2019,Cuenca2019}, and even remote surgery \cite{Laaki2019Prototyping} and big data \cite{Khatib2016}.
Despite of the advantages, networked control systems also pose challenging problems arising from
unreliability in data transmission, caused by channel induced delays (i.e., data transmission times), packet
dropout, coding error etc.
Due to the fact that data are transmitted in network by data packet, channel induced delays and data packet
dropout that occur in data exchanging between components of feedback systems over network are the most
common phenomena in networked systems. The channel induced delays and packet dropout could
seriously degrade performance of feedback control systems, and even destabilize feedback systems when
these issues are not carefully considered in system design (see, e.g., \cite{Hespanha2007, Park2018} and \cite{ZHANG2017Survey}). 
To cope with channel induced delays, great efforts have been made in modeling, stability analysis, and
controller design of networked feedback systems (for example see \cite{Hespanha2007, Park2018} and the
references therein).

Channel induced delays in networked systems could be constant, time-varying\hltb{, and} random delays,
which are basically dependent {on} network protocol, channel quality etc. In \cite{Branicky2000, Walsh1999,
Zhang2001Stability}, the authors studied the {asymptotic} stability of networked linear feedback systems in which message is transmitted over network with a constant transmission time (i.e. a constant channel {induced} delay). The networked systems were modeled as discrete-time linear systems augmented from discretized plants, delay channel models and control laws. In particular, the channel induced delay was {modeled} as a parameter of the augmented linear {systems' matrices}.  Stability criteria were obtained for the networked systems in terms of the discrete-time linear system models. These augmented models were also used to study the {asymptotic} stability and stabilization design for networked feedback systems with time-varying delays. Several stability criteria and stabilization design results were obtained by using switch system approaches (see for example \cite{Lin2005Stability, Lin2009Survey, ZHANG20083206, Zhang2016Survey}).
Alternatively, time-varying channel induced delays were modeled as continuous-time plants' input delays
(see e.g., \cite{FridmanSS_SCL2005, Kharitonov200315, Kharitonov2003}).  Lyapunov-Krasovskii functionals
were used in stability analysis and stabilization design for networked feedback systems. Various linear matrix
inequality (LMI) based methods were {developed}. Moreover, to cope with data disorder {occurred} in transmission, time-stamp scheme and logic \hltb{zero-order hold} were developed for the networked feedback systems \cite{NILSSON199857, Xiong2009}. In general, the aims of aforementioned papers are to find
criteria that networked feedback systems are stable when channel induced delays belong to given regions,
or to find upper bounds of channel induced delays under which the stability of the networked feedback
systems are preserved. This leads to certain conservativeness in stability analyses and stabilization design
for networked feedback systems with random channel induced delays.

Since channel induced delays and packet dropouts usually exhibit random characteristics, random channel
induced delay and packet dropout models are widely used in stability analysis and control design for
networked feedback systems (see for example \cite{DONKERS2012917, NILSSON199857, QuevedoJ_TAC2014, Xiong2007Stabilization, Zhang2005A}).
In \cite{Xiong2007Stabilization, Zhang2005A}, applying {Markov chains},
the authors modeled networked systems with random delays and packet dropouts as discrete-time Markov
jump linear systems. Necessary and sufficient conditions were obtained for the mean-square stability of the
networked feedback systems. Then, the mean-square stabilization was studied for these systems and LMI
based control design {approaches} were developed. \hltb{It turned out} that this stabilization problem is a non-convex problem. Certain conservativeness may not be {avoidable} in this type of results. On the other hand, channel induced delays were modeled as random input delays of plant/controllers, and the sequences
of the input delays were assumed to be independent and identically distributed (i.i.d.) random processes in
\cite{DONKERS2012917,  NILSSON199857}. It was assumed in these works that the probability density
functions of the random channel induced delays are known and all the channel induced delays in
networked feedback systems are smaller than the systems' sampling intervals.
A suboptimal linear controller was designed for a networked feedback system with a fixed sampling interval in \cite{NILSSON199857}.
The reference \cite{DONKERS2012917} studied a networked system with a time-varying sampling interval.
LMI based sufficient conditions were presented for the mean-square stability of the system.
However, for the case when channel induced delays are longer than the systems' sampling intervals, the channel induced delays may not exactly be input delays of plants since {more than one} data may be received simultaneously or none data could be received at one sampling instant.
In \cite{Chen2014Distributed, QuevedoJ_TAC2014,Schenato2008Optimal, Shi2010Kalman}, channel induced delays (or data transmission times) which are longer than the systems' sampling intervals were modeled as an i.i.d. random process with a known
probability mass function (PMF). State estimation problem was studied for these networked systems and
optimal state estimation algorithms were developed in \cite{ Schenato2008Optimal, Shi2010Kalman}. The stabilization problem was studied for a nonlinear networked feedback system and a sufficient condition was obtained for the stability of the networked feedback system in \cite{QuevedoJ_TAC2014}.

In this work, we focus on a single-input single-output linear time-invariant (LTI) networked feedback system in which the control signal is transmitted
over a communication channel with random channel induced delays.
The networked feedback system is a discrete-time system with a fixed sampling interval and channel
induced delays are integral multiples of this sampling interval.
Our goal is to fully understand how the interaction between stochastic features of the random delays and characteristics of the networked feedback system affects the system's stability and limits the stabilizability of the system.
With these purposes, we adopt the random process model studied in \cite{Chen2014Distributed, QuevedoJ_TAC2014, Schenato2008Optimal, Shi2010Kalman} to describe the random channel induced delays.
\hlt{At the channel terminal, a linear receiving strategy which is a linear combination with given weights of data received at one sampling time
is adopted. The new signal generated by the linear receiving strategy is applied to the plant of the networked feedback system as a control signal.}
Then, the channel {uncertainty} caused by the random channel induced delays is defined by the random impulse response of the communication channel and the first-order statistics of the impulse response.
The input-output relation of the channel uncertainty is established in terms of the spectral density of the \hltb{uncertainty's} impulse response. To give a precise description for a channel relative deviation induced by the channel uncertainty, a concept \hltb{of so-called} {\em {frequency response of variation}} is introduced in frequency domain. A necessary and sufficient condition is presented for the \msis of the networked feedback system. It is a new version of the small gain theorem for networked feedback systems with random channel induced delays. {{It is also found that for the \msis and stabilizability problems, the channel uncertainty caused by the random channel induced delays is equivalent to an additive noise in an additive noise channel with a signal-to-noise ratio constraint (see  \cite{Franceschetti2012} and \cite{GonzalezChen2019} \hltb{for more details} in the mean-square stability of networked feedback systems over additive noise channels with a signal-to-noise ratio constraint)}.
After then, the mean-square input-output stabilizability via output feedback is studied for the networked feedback system. In particular, a necessary and sufficient condition is found for the mean-square input-output stabilizability of the system when its plant is minimum phase. It precisely describes the connection between the mean-square input-output stabilizability, the {{frequency response of variation}} and 
the unstable poles of the plant in the feedback system.
Furthermore, \hltb{it turns} out that the interaction between the {{frequency response of variation}} and the
unstable poles plays a critical role to the mean-square stabilizability of the networked feedback system.

The remainder of this paper is organized as follows. In Section \ref{Sec:Problem_Formulation}, the random
channel induced delays and related channel uncertainty are modeled by using the PMF of the delays.
The impulse responses of the channel and channel uncertainty are given.
The problems under study are formulated. Section \ref{Sec:Norm_condition} presents the input-output relation
of the channel uncertainty based on the spectral density of its impulse response.
A necessary and sufficient condition of the \msis is obtained for the networked feedback system. In Section \ref{FSNR_UP}, the mean-square input-output stabilizability is studied for the networked feedback
system when its plant is minimum phase. An analytic expression is obtained for the necessary and sufficient condition of the mean-square input-output stabilizability. The connection between the result in this work and
the existing results is discussed. Section \ref{Sec:Examples} illustrates some numerical examples and
Section \ref{Sec:Conclusion} concludes the paper.

The notations used in this paper is mostly standard.
The complex conjugate transpose of any matrix $A$ is denoted by $A^*$.
When $A$ is square and invertible, its inverse and inverse conjugate transpose are denoted by $A^{-1}$ and $A^{-*}$, \hltb{respectively}.
For any transfer function $G(z)$, we represent a state-space realization of $G(z)$ by
$G(z)=\left[\begin{array}{c|c}
 A & B \\ \hline  C & D \end{array}\right]$.
Let the open unit disc be denoted by $\mathbb{D} := \{z \in \mathbb{C}:|z|<1\}$, the closed unit disc by
$\bar{\mathbb{D}} := \{z \in \bar{\mathbb{C}}:|z|\le1\}$, the unit circle by $\partial \mathbb{D}$, and the
complements of $\mathbb{D}$ and $\bar{\mathbb{D}}$ by $\mathbb{D}^c$ and $\bar{\mathbb{D}}^c$, respectively.
In this work, the Hardy space $\mathcal{H}_2$ consists of scalar-valued analytic function $F$ in
$\bar{\mathbb{D}}^c$ such that
\begin{equation*}
\|F\|_2 =
\Big(\sup_{r>1}\frac{1}{2\pi}\int_{-\pi}^{\pi}F^*(re^{j\theta})F(re^{j\theta})d\theta\Big)^{\frac{1}{2}}<\infty.
\end{equation*}
The orthogonal complement of $\mathcal{H}_2$ is given by
\begin{equation*}
\begin{aligned}
&\mathcal{H}_2^\perp := \Big\{F:F(z)~\text{analytic in}~\mathbb{D},~F(0)=0,\\
&\hspace{2.5cm}\|F\|_2 = \Big(\sup_{r <
1}\frac{1}{2\pi}\int_{-\pi}^{\pi}F^*(re^{j\theta})F(re^{j\theta})d\theta\Big)^{\frac{1}{2}}<\infty \Big\}.
\end{aligned}
\end{equation*}
Define also the space $\mathcal{RH}_\infty$ as the set of all proper stable rational transfer functions.
Furthermore, $\E\{\cdot\}$ denotes the expectation operator of a random variable. The set of real numbers
is denoted by $\mathbb{R}$.

\section{Problem Formulation}\label{Sec:Problem_Formulation}

Consider a canonical structure of a discrete-time LTI networked feedback system as depicted
in Fig. \ref{Fig:System_model}. Here, the plant $P$ is an LTI system and its transfer function
$P(z)$ is assumed to be strictly proper. The controller $K$ is an {LTI} controller.
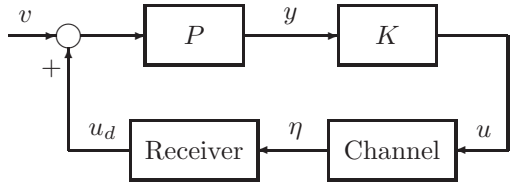
\begin{figure}[hbt]
  \centering
\setlength{\unitlength}{.25in}
\linethickness{.75pt}
\begin{picture}(9,3.5)
\put(-0.1,3){\circle{.5}}

\put(0.15,3){\vector(1,0){1.35}}
\put(1.5,2.4){\framebox(2,1.2){$P$}}
\put(4.5,3.4){\makebox(0,0){$y$}}
\put(3.5,3){\vector(1,0){2}}
\put(5.5,2.4){\framebox(2,1.2){$K$}}
\put(7.5,3){\line(1,0){1.5}}

\put(9,3){\line(0,-1){2.4}}
\put(9,0.6){\vector(-1,0){1.1}}
\put(8.5,1){\makebox(0,0){$u$}}
\put(5.3,0){\framebox(2.6,1.2){Channel}}
\put(4.6,1){\makebox(0,0){${\bf \eta}$}}
\put(0.6,1){\makebox(0,0){$u_d$}}
\put(1.2,0.6){\line(-1,0){1.3}}
\put(1.2,0){\framebox(2.6,1.2){Receiver}}
\put(5.3,0.6){\vector(-1,0){1.5}}
\put(-0.1,0.6){\vector(0,1){2.15}}
\put(-0.45,2.3){\makebox(0,0){$+$}}

\put(-1.35,3){\vector(1,0){1}}
\put(-1,3.4){\makebox(0,0){$v$}}
\end{picture}
\caption{Networked feedback system over an unreliable communication channel}
\label{Fig:System_model}
\end{figure}

The signal $y(k) \in \mathbb{R}$ is the measurement of the plant $P$,
$u(k)\in \mathbb{R}$ is the control signal generated by the linear controller $K$, and $v(k)$ is the external input.
An unreliable communication channel is placed in the path from the controller to the receiver.
Here, the {unreliable} features under study are random channel induced delays and packet dropout.
Denote the channel induced delay for the control signal $u(k)$ (i.e., the transmission time {spent on}
transmitting $u(k)$ over the {communication} channel) by $\tau_k$.
Thus, the signal $u(k)$ sent at the time $k$ arrives at its destination at the time
$k+\tau_k$. The channel induced delay $\tau_k$ is
assumed to be a random variable with nonnegative integer values from a bounded set
$\mathcal{D}=\{0, 1, 2, \cdots, \bar{\tau}-1, \bar{\tau}\}$.
All transmitted data are with time stamps. The data whose channel induced delays are greater than $\bar{\tau}$ are discarded at the receiver.
${\bf \eta}(k)$ is a collection of data, which has $\bar{\tau}+1$ entries and includes all data received at time $k$.
Now, we use Kronecker delta function $\delta(k)$ to characterize all entries belonging to ${\bf \eta}(k)$,
\begin{align*}
\delta(k)=\left\{ \begin{matrix}  1, & k=0;  \\  0, &k\neq 0.   \end{matrix}     \right.
\end{align*}
Let ${\bf \eta}(k)=\left\{\eta_0(k), \eta_1(k), \cdots, \eta_{\bar{\tau}}(k)\right\}$ and
$\eta_i(k) = \delta(\tau_{k-i}-i)u(k-i),~i\in \mathcal{D}$. The function  $\delta(\tau_{k-i}-i) = 1$ indicates that
the signal $u(k-i)$ arrives at its destination through the channel at time $k$ and $\eta_i(k)=u(k-i)$,  otherwise it means that $u(k-i)$ is not received at time $k$ and $\eta_i(k)=0$. It is assumed that
the receiver at the terminal of the channel has limited computation capability. This capability allows the receiver to generate the
signal $u_d(k) \in \mathbb{R}$ as its output based on
the received data and the time stamps associated with the data.
A linear combination of the received data
$\{\eta_0(k),\cdots,\eta_{\bar{\tau}}(k)\}$ can be taken as the output of the receiver:
\begin{align}\label{Equ:delay_model}
u_d(k)= \sum_{i=0}^{\bar{\tau}} \alpha_i\eta_i(k)
=\sum_{i=0}^{\bar{\tau}} \alpha_i\delta(\tau_{k-i}-i)u(k-i)
\end{align}
where the weights $\alpha_0,\alpha_1,\ldots,\alpha_{\bar{\tau}}$ are assigned to the received data,
respectively, according to the delay steps of the data.
\hlt{We refer to the receiver given by \eqref{Equ:delay_model} as linear receiver or \emph{linear receiving strategy}.
Note that the linear receiving strategy is a general case of the zero-input strategy \cite{Schenato2009ZeroHold}.}
Without loss of generality, we assume that the initial time is at $k=0$ and the system is at rest at initial time. 
\begin{remark}
For any given $k$ and a realization of $\tau_k$, the indicator $\delta(\tau_{k}-i)$ guarantees that the control
signal $u(k)$ could only appear in one {element} of the sequence $\left\{{\bf \eta}(0), {\bf \eta}(1), {\bf \eta}(2), \cdots \right\}$.
Being consistent with ${\bf \eta}(k)$, the data $u(k)$ appears in the sequence $\left\{ u_d(0), u_d(1), u_d(2), \cdots \right\}$
once at most. Moreover, to drop data with a channel induced delay greater than $\bar{\tau}-1$ and model data dropout, the weight $\alpha_{\bar{\tau}}$ is set to be zero.
{That is, a zero-input strategy is adopted.}
\end{remark}

According to the discussion above, the networked feedback system in Fig. \ref{Fig:System_model} \hlt{with the linear receiver} is re-diagrammed as that shown in Fig. \ref{Fig:System_connected_channel}, wherein $\q^{-1}$ is the unit delay operator.
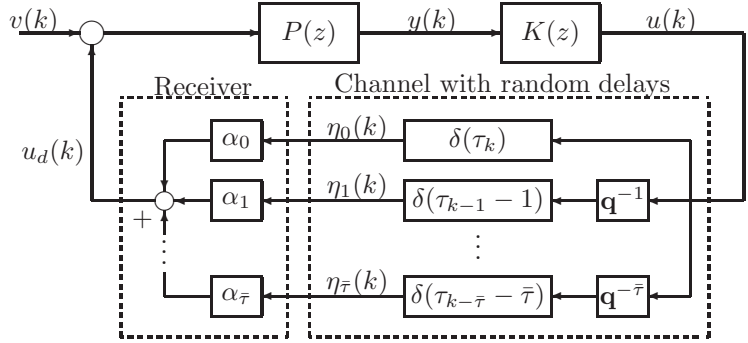
\begin{figure}[!hbt]
  \centering
\setlength{\unitlength}{.25in}
\linethickness{.75pt}
\begin{picture}(13.5,7)
\put(0,6.5){\circle{0.5}}
\put(-1.5,6.5){\vector(1,0){1.25}}
\put(-1.2,6.75){\makebox(0,0){$v(k)$}}

\put(0.25,6.5){\vector(1,0){3.25}}
\put(3.5,5.9){\framebox(2,1.2){$P(z)$}}
\put(5.5,6.5){\vector(1,0){3}}
\put(7.05,6.75){\makebox(0,0){$y(k)$}}
\put(8.5,5.9){\framebox(2,1.2){$K(z)$}}
\put(10.5,6.5){\line(1,0){3}}
\put(12,6.75){\makebox(0,0){$u(k)$}}
\put(13.5,6.5){\line(0,-1){3.5}}

\put(8,2.2){\makebox(0,0){$\vdots$}}

\put(-0.8,4){\makebox(0,0){$u_d(k)$}}

\put(13.5,3){\vector(-1,0){2}}
\put(12.4,3){\line(0,1){1.25}}
\put(12.4,4.25){\vector(-1,0){2.9}}
\put(12.4,3){\line(0,-1){2}}
\put(12.4,1){\vector(-1,0){0.9}}

\put(6.5,3.85){\framebox(3,0.8){$\delta(\tau_{k})$}}
\put(5.5,4.5){\makebox(0,0){$\eta_0(k)$}}
\put(6.5,4.25){\vector(-1,0){3}}
\put(2.5,3.85){\framebox(1,0.8){$\alpha_0$}}
\put(2.5,4.25){\line(-1,0){1}}
\put(1.5,4.25){\vector(0,-1){1.05}}

\put(10.5,2.6){\framebox(1,0.8){${\q}^{-1}$}}
\put(10.5,3){\vector(-1,0){1}}
\put(6.5,2.6){\framebox(3,0.8){$\delta(\tau_{k-1}-1)$}}
\put(5.5,3.3){\makebox(0,0){$\eta_1(k)$}}
\put(6.5,3){\vector(-1,0){3}}
\put(2.5,2.6){\framebox(1,0.8){$\alpha_1$}}
\put(2.5,3){\vector(-1,0){0.8}}
\put(1.5,3){\circle{0.4}}

\put(10.5,0.6){\framebox(1,0.8){$\q^{-\bar{\tau}}$}}
\put(10.5,1){\vector(-1,0){1}}
\put(6.5,0.6){\framebox(3,0.8){$\delta(\tau_{k-\bar{\tau}}-\bar{\tau})$}}
\put(5.5,1.3){\makebox(0,0){$\eta_{\bar{\tau}}(k)$}}
\put(6.5,1){\vector(-1,0){3}}
\put(2.5,0.6){\framebox(1,0.8){$\alpha_{\bar{\tau}}$}}
\put(2.5,1){\line(-1,0){1}}
\put(1.5,1){\line(0,1){0.5}}
\put(1.5,2.1){\makebox(0,0){$\vdots$}}

\put(1.5,2.3){\vector(0,1){0.5}}
\put(1.05,2.6){\makebox(0,0){$+$}}

\put(1.3,3){\line(-1,0){1.3}}
\put(0,3){\vector(0,1){3.25}}

\put(8.5,5.4){\makebox(0,0){Channel with random delays}}
\put(4.5,0.15){\dashbox{0.1}(8.28,5){}}

\put(2.3,5.4){\makebox(0,0){Receiver}}
\put(0.6,0.15){\dashbox{0.1}(3.45,5){}}
\end{picture}
  \caption{Networked feedback system with a random delay channel}\label{Fig:System_connected_channel}
\end{figure}
One can see from this block-diagram that the block cascaded by the channel and the receiver, referred to
as a transmission block, is a linear system with a random finite impulse response. Its input-output relation
is given by (\ref{Equ:delay_model}). Since the initial time is assumed to be zero, without loss generality, we consider the random impulse response $h(k, n)$ of the transmission block to a
unit impulse input applied to the channel at any time $n \geq 0$, which is given by
\begin{equation}\label{R_FIR}
h(k,n)=\left\{\begin{array}{lll}  0,  & -\infty < k< n,   \\
                              \alpha_i \delta(\tau_n-i),  & k=n+i, i \in \mathcal{D},  \\
                              0,  & k > n+\bar{\tau}. \end{array}\right.
\end{equation}
One can see from (\ref{R_FIR}) that the random impulse response $h(k,n)$ is determined by the instant $n$ and the random sequence $\{\alpha_0\delta(\tau_n), \alpha_1\delta(\tau_n-1), \cdots, \alpha_{\bar{\tau}}\delta(\tau_n-\bar{\tau})\}$. For a given realization of $\tau_n$, at most one entry is not equal to zero in the random sequence.
The input-output relation of the transmission block is rewritten as
\begin{align*}
  u_d(k) &= \sum_{i=0}^{\bar{\tau}} h(k,k-i) u(k-i).
\end{align*}

We impose the following assumptions throughout the paper.
\begin{assum}\label{Assump:Sum_of_probability}
The random delay process $ \{\tau_k: k=0,1,2, \cdots  \}$ is an i.i.d. process, and $\tau_k$ takes values in
$\mathcal{D}$ according to a common PMF that
\begin{equation}\label{Equ:Prob_Mass_Func}
p_i=\Pr \left\{ {{\tau_k} = i} \right\},~~i \in \mathcal{D}
\end{equation}
with $p_i \in [0,1]$ and $\sum_{i \in \mathcal{D}} p_i  = 1$.
\end{assum}

{
\begin{assum} \label{Assump:independent_assumption}
The external input sequence $\{v(0),v(1),v(2),\cdots\}$ is independent of the channel induced delay process $\{\tau_0,\tau_1,\tau_2,\cdots\}$.
\end{assum}

}

Since the random sequence $\{\alpha_0\delta(\tau_n), \alpha_1\delta(\tau_n-1), \cdots, \alpha_{\bar{\tau}}\delta(\tau_n-\bar{\tau})\}$ is dependent on the random variable $\tau_n$ which is an entry of an i.i.d. process,  the mean of each entry in this sequence is obtained from the PMF of $\tau_n$, i.e.,
\begin{align}\label{mean_channel_FIR}
{\E}\{\delta(\tau_n-i)\}=p_i, \quad i \in \mathcal{D}.
\end{align}
Define the {\emph{mean channel}} as
\begin{align*}
H(z) = \sum \limits_{i=0}^{\bar{\tau}}\alpha_i {\E}\{\delta(\tau_n-i)\} z^{-i}.
\end{align*}
Thus, it holds that
\begin{align}\label{Equ:H}
H(z) = \sum \limits_{i=0}^{\bar{\tau}}\alpha_i p_i z^{-i}.
\end{align}

Subsequently, the transmission block is divided into two parts: One is the mean channel $H(z)$ and the other is a
zero-mean channel uncertainty denoted by $\Omega$. Denote the response of the latter part by $\omega(k, n)$ to
the unit impulse input applied {to} the channel at time $n \geq 0$. This impulse response is given as below:
\begin{equation}\label{Equ:Omega_response_1}
\omega(k, n)= \left\{ \begin{array}{ll}
						0, & k<n,\\
                         \alpha_i[\delta(\tau_{n}-i)-p_i],& k=n+i, i \in {\mathcal D}, \\
                          0,&  k>n+\bar{\tau}.
                        \end{array} \right.
\end{equation}
Accordingly, the receiver output $u_d(k)$ is the summation of the outputs of $H$ and $\Omega$ when considering
$u(k)$ as their inputs, i.e.,
\begin{align}\label{Equ:Channel_output}
u_d(k)={\bar{u}}(k)+d(k)
\end{align}
where
\begin{align}
{{\bar{u}}}(k) &= \sum_{i = 0}^{\bar{\tau}} \alpha_i p_i u(k-i),\label{Equ:u_hat}\\
d(k) &= \sum_{i = 0}^{\bar{\tau}} \omega(k, k-i) u(k-i). \label{Equ:d_total}
\end{align}

As a result, the system in Fig. \ref{Fig:System_connected_channel} can be re-diagramed as a stochastic system
shown in Fig. \ref{Fig:System_G_with_uncertainty}. The structure of this system is similar to that studied in
literatures for networked feedback systems over fading channels (see for example \cite{Elia2005Remote} and
\cite{Elia2011limitations}). In the literatures, the channel uncertainties under study are white noise processes,
thus the mean channels are constants and the channel uncertainties are zero-mean white noise processes. But, in
this work, the mean channel $H(z)$ and channel uncertainty $\Omega$ are linear systems with \hlt{ the time-invariant finite
impulse response $\left\{\alpha_0 p_0, \cdots, \alpha_{\bar{\tau}} p_{\bar{\tau}}  \right\}$ and the random finite impulse response  $\left\{\omega(n, n), \cdots, \omega(n+\bar{\tau}, n) \right\}$, respectively.}

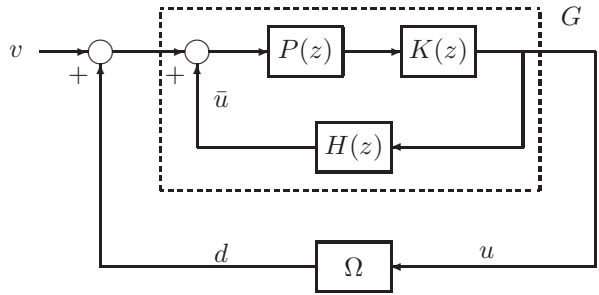
\begin{figure}[!hbt]
  \centering
\setlength{\unitlength}{.25in}
\linethickness{.75pt}
\begin{picture}(12,6.1)
\put(1.25,5){\circle{0.5}}
\put(0,5){\vector(1,0){1}}
\put(-0.5,5){\makebox(0,0){$v$}}
\put(1.5,5){\vector(1,0){1.5}}
\put(3.25,5){\circle{.5}}
\put(3.5,5){\vector(1,0){1.25}}
\put(4.75,4.5){\framebox(1.5,1){$P(z)$}}
\put(6.25,5){\vector(1,0){1.25}}
\put(7.5,4.5){\framebox(1.5,1){$K(z)$}}
\put(9,5){\line(1,0){2.5}}
\put(10,5){\line(0,-1){2}}
\put(11.5,5){\line(0,-1){4.5}}
\put(11,5.75){\makebox(0,0){$G$}}
\put(10,3){\vector(-1,0){2.75}}
\put(5.75,2.5){\framebox(1.5,1){$H(z)$}}
\put(5.75,3){\line(-1,0){2.5}}
\put(3.25,3){\vector(0,1){1.75}}
\put(2.8,4.5){\makebox(0,0){$+$}}
\put(3.75,4){\makebox(0,0){$\bar{u}$}}

\put(11.5,0.5){\vector(-1,0){4.25}}

\put(9.25,0.8){\makebox(0,0){$u$}}
\put(5.75,0){\framebox(1.5,1){$\Omega$}}
\put(3.75,0.8){\makebox(0,0){$d$}}

\put(5.75,0.5){\line(-1,0){4.5}}
\put(1.25,0.5){\vector(0,1){4.25}}
\put(0.8,4.5){\makebox(0,0){$+$}}

\put(2.5,2.1){\dashbox{0.1}(7.85,3.8){}}
\end{picture}
  \caption{The stochastic system interconnected by a nominal system and a channel uncertainty
  }\label{Fig:System_G_with_uncertainty}
\end{figure}

In Fig. \ref{Fig:System_G_with_uncertainty}, denote the system from $d$ to $u$ without considering the channel
uncertainty $\Omega$ by $G(z)$, referred to as the {\emph{nominal system}}, which is given by
\begin{equation}\label{Equ:Nominal_system_G(z)}
G(z) = {K(z)P(z)}[{1 - H(z)K(z) P(z)}]^{-1}.
\end{equation}
Then the whole system is an interconnection of the nominal system $G(z)$ and the channel uncertainty $\Omega$, as
shown in Fig. \ref{Fig:G_Omega}.

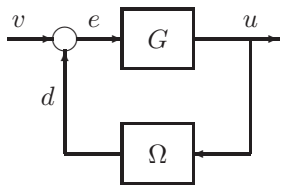
\begin{figure}[!hbt]
\centering
\setlength{\unitlength}{.3in}
\linethickness{.75pt}
\begin{picture}(5.2,3)
\put(0.2,2.8){\makebox(0,0){$v$}}
\put(1.5,2.8){\makebox(0,0){$e$}}
\put(0,2.5){\vector(1,0){0.8}}
\put(1,2.5){\circle{0.4}}
\put(1.2,2.5){\vector(1,0){0.8}}
\put(2,2){\framebox(1.2,1){$G$}}
\put(3.2,2.5){\vector(1,0){1.5}}
\put(4.2,2.8){\makebox(0,0){$u$}}
\put(4.2,2.5){\line(0,-1){2}}
\put(4.2,0.5){\vector(-1,0){1}}
\put(2,0){\framebox(1.2,1){$\Omega$}}
\put(2,0.5){\line(-1,0){1}}
\put(0.7,1.5){\makebox(0,0){$d$}}
\put(1,0.5){\vector(0,1){1.8}}
\end{picture}
\caption{Equivalent interconnection of a networked feedback system in presence of random channel induced delays. 
}\label{Fig:G_Omega}
\end{figure}

It is well-known that there exists a controller $K$ to internally stabilize the nominal feedback system $G$ for any stabilizable and detectable LTI plant $P$ if and only if there is not unstable pole-zero cancelation between the mean channel $H$ and the plant $P$.
That is, the following assumption is necessary.

\begin{assum}\label{Assump:K_stabilizes_G}
{The plant $P$ of the networked feedback system is stabilizable and detectable.}
There is not unstable pole-zero cancelation between the mean channel $H$ and the plant $P$.
\end{assum}

To avoid any possible unstable pole-zero cancelation, the weights $\alpha_0, \alpha_1$, $\cdots, \alpha_{\bar{\tau}}$ \hlt{in the linear receiving strategy} can also be selected so that the mean channel $H$ is minimum phase.
A numerical example is presented in Section \ref{Sec:Examples}.
\hlt{Since the controller-receiver co-design is a very difficult task in general,
a common setup in most of literatures is that the controller is designed based on a given receiving strategy, e.g., zero-input strategy or hold-input strategy for packet dropout problem \cite{Elia2005Remote,Schenato2009ZeroHold}.
In this work, we restrict ourselves to the fixed weights $\{\alpha_0,\cdots, \alpha_{\bar{\tau}}\}$ of the linear receiving strategy such that $H$ is minimum phase.}
Throughout this paper, we concentrate on the \msis and stabilizability defined next.

\begin{defn}\label{Def:Internally_MS}
The channel induced delay process $\{\tau_k, k=0,1,2, \cdots\}$ satisfies Assumption \ref{Assump:Sum_of_probability}.
The networked feedback system \hlt{with a given linear receiving strategy} shown in Fig. \ref{Fig:System_connected_channel} is mean-square \hlt{input-output} stable if the linear controller $K(z)$ internally stabilizes $G(z)$ and 
the signal sequence $\{u(k)\}$ is with bounded variances for any i.i.d. input process {$\{v(k)\}$} with a bounded variance and independent of the channel induced delay process.
\end{defn}

Let the set of all proper controllers internally stabilizing $G(z)$ be $\mathcal{K}$.

\begin{defn}\label{Def:Mean_square_stabilizable}
The networked feedback system \hlt{with a given linear receiving strategy} shown in Fig. \ref{Fig:System_connected_channel} is said to be \hlt{mean-square input-output stabilizable} via output feedback if there exists a feedback controller $K \in \mathcal{K}$ such that the
closed-loop system is mean-square \hlt{input-output} stable.
\end{defn}

\begin{remark}
For a memoryless channel with a memoryless uncertainty, the channel model is given by
\begin{equation}\label{Equ:Typical_channel}
u_d(k) = (\mu+\Delta_k)u(k)
\end{equation}
where $\mu$ is the mean of the channel gain and $\{\Delta_k\}_{k=0}^\infty$ is an i.i.d. process with zero mean and finite variance $\sigma^2$.
It is well-known (for example see \cite{Elia2005Remote,Qi2017Control}) that the networked feedback system over an
unreliable channel modeled by \eqref{Equ:Typical_channel} is mean-square (input-output) stable if and only if $ KP (1-\mu KP)^{-1}$ is internally stable and
\begin{equation}\label{Equ:Classical_mean-square-stability}
\Big\| \frac{\sigma}{\mu}[\mu KP (1-\mu KP)^{-1}]\Big\|_2^2 < 1,
\end{equation}
{where $\sigma/\mu$ is referred to as the {\em relative standard deviation} or {\em coefficient of variation} (see for example \cite{Stewart2009Probability}) of the random variable $\mu+\Delta_k$. It is a measure to the variation of the uncertain  channel's gain.
In this paper, since the channel uncertainty under study is with memory, it is much more complicated than that induced by an i.i.d multiplicative noise. A generalized version of the \msis criterion is studied for the networked feedback system.}
\end{remark}

\section{Mean-square Input-output Stability}\label{Sec:Norm_condition}

In this section, we study the criterion of \msis for the networked feedback system with a given output feedback controller $K \in {\cal K}$.
The frequency variable $z$ will be omitted whenever no confusion is caused.

It is shown in preceding section that a networked feedback system with a random channel induced delay \hlt{and a linear receiver} is modeled as a stochastic system shown in Fig. \ref{Fig:System_G_with_uncertainty}.
Intuitively, the \msis of the system is determined by the interaction between the nominal feedback system $G$ and the zero-mean channel uncertainty $\Omega$.
To study the interaction between $G$ and $\Omega$, the stochastic properties of the channel uncertainty $\Omega$ whose impulse response is given in (\ref{Equ:Omega_response_1}) are studied.

\begin{lem}\label{Lma:omega_i}
Suppose that the random delay process $\{\tau_k:k=0,1,2,\cdots\}$ satisfies Assumption
\ref{Assump:Sum_of_probability}. Then it holds to the impulse response of the channel uncertainty
$\Omega$ that
\begin{enumerate}
\item
for $i \in \mathcal{D}$ and $k=0,1,2,\cdots$,
\begin{equation*}
\E\{\omega(k,k-i)\} = 0;
\end{equation*}
\item for $i \in \mathcal{D}$, $k_1,k_2=0,1,2,\cdots,$
\begin{equation*}
\hspace{-0.3cm}
\E\left\{ \omega(k_1, k_1-i) \omega(k_2, k_2-i) \right\}= \delta (k_1 - k_2) \alpha_i^2 p_i{(1 - p_i)};
\end{equation*}
\item for $i_1 \neq i_2$, $i_1,i_2 \in \mathcal{D}$, $k_1, k_2=0,1,2,\cdots$,
\begin{align*}
&\E\left\{ \omega(k_1,k_1-i_1) \omega(k_2, k_2-i_2) \right\}
= - \delta (k_1-i_1-k_2+i_2) \alpha_{i_1} \alpha_{i_2} p_{i_1} p_{i_2}.
\end{align*}
\end{enumerate}
\end{lem}
\begin{proof}
It straightforwardly follows from the impulse response of $\Omega$ in  (\ref{Equ:Omega_response_1})
that Lemma \ref{Lma:omega_i}.1 holds.

From Assumption \ref{Assump:Sum_of_probability}, $\tau_{k_1-i}$ is independent of $\tau_{k_2-i}$ for any $k_1 \neq k_2$. Thus, it holds that
\begin{align*}
\E\left\{ \omega(k_1, k_1-i) \omega(k_2, k_2-i) \right\}
=\E\left\{ \omega(k_1, k_1-i)\right\} \E\left\{\omega(k_2, k_2-i) \right\}=0.
\end{align*}
According to the PMF of $\tau_{k_1-i}$ in (\ref{Equ:Prob_Mass_Func}), it holds for any $k_1 = k_2$ that
\begin{align*}
\E\left\{ \omega(k_1, k_1-i) \omega(k_2, k_2-i) \right\}
=\alpha_i^2[(1-p_i)^2p_i+p_i^2(1-p_i)]=\alpha_i^2 p_i(1-p_i).
\end{align*}
Hence, Lemma \ref{Lma:omega_i}.2 holds.

Now consider Lemma \ref{Lma:omega_i}.3. It also follows from Assumption \ref{Assump:Sum_of_probability} that for any $k_1-i_1\neq k_2-i_2$, $\tau_{k_1-i_1}$ is independent of $\tau_{k_2-i_2}$. It yields that
\begin{align*}
\E\left\{ \omega(k_1, k_1-i_1) \omega(k_2, k_2-i_2) \right\}
=\E\left\{ \omega(k_1, k_1-i_1)\right\} \E\left\{\omega(k_2, k_2-i_2) \right\}=0.
\end{align*}
In the case $k_1-i_1 = k_2-i_2$, according to the PMF of $\tau_{k_1-i_1}$ in (\ref{Equ:Prob_Mass_Func}),
we have that
\begin{align*}
&\E\left\{ \omega(k_1, k_1-i_1) \omega(k_2, k_2-i_2) \right\}\\
=&\alpha_{i_1}\alpha_{i_2} [(1 - p_{i_1})(-p_{i_2})p_{i_1} -p_{i_1}(1-p_{i_2})p_{i_2}
+ p_{i_1} p_{i_2} (1-p_{i_1}- p_{i_2})]\\
=&- \alpha_{i_1}\alpha_{i_2}p_{i_1}p_{i_2}.
\end{align*}
This completes the proof.
\end{proof}

\hlt{
\begin{remark}
According to the definition of $\omega(k,n)$ in (\ref{Equ:Omega_response_1}),
it holds that $\omega(k,n) \equiv 0$ for $k<n$ and $k > n+\bar{\tau}$, i.e., there are only $(1+\bar{\tau})$ non-zero elements in the response
$\left\{\omega(k,n): -\infty <k<\infty \right\}$ of $\Omega$ to the impulse input
$\delta(k-n)$.
Thus, $\left\{\omega(k,k-i):k\geq 0,\; i \in {\cal D} \right\}$ is the collection of all non-zero elements in the sequence
$\left\{\omega(k,n): -\infty <k<\infty,\; n\geq -{\bar{\tau}} \right\}$.
Furthermore, it holds that
\begin{align*}
\left\{\omega(k,k-i): k\geq 0,\; i \in {\cal D} \right\}
=\left\{\omega(n+i, n):  n\geq -\bar{\tau}, \; i \in {\cal D} \right\}.
\end{align*}
Lemma \ref{Lma:omega_i} shows that for any given $n \geq -\bar{\tau}$, the first- and second-order statistics of all elements in the non-zero subsequence
$\{\omega(n+i,n): i \in {\cal D} \}$ are determined by the PMF of the transmission time $\tau_n$
and independent of $n$. For $n_1 \ne n_2$ ($n_1, n_2 \geq -\bar{\tau}$), the subsequences
$\{\omega(k,n_1)\}$ and $\{\omega(k,n_2)\}$ are mutually independent.
From the proof of this lemma, it can be verified that these stochastic properties
holds for all $-\infty<n<\infty$.
\end{remark}
}

Now the second order stochastic properties of $\Omega$ are studied.
For any given instant $n \geq -\bar{\tau} $, \hlt{let the autocorrelation of the subsequence $\{\omega(k,n): -\infty< k< \infty\}$ be given by }
\begin{equation}\label{Equ:Relative}
\begin{aligned}
r(l)= {\E} \Bigg\{\sum_{k = -\infty}^{\infty} \omega(k,n)\omega(k+l,n)\Bigg\}, \quad
-\infty < l < \infty.
\end{aligned}
\end{equation}

Note the fact that $\omega(k, n) \equiv 0$ for $k<n$ and $k> n+\bar{\tau}$. 
}
{For the case $l=0$, only the terms with $n \leq k \leq n+\bar{\tau}$ in the summation of (\ref{Equ:Relative}) may not be equal to zero.
By letting $k_1=k_2=k$, $i=k-n$ and applying Lemma \ref{Lma:omega_i}.2 into \eqref{Equ:Relative}, we obtain that
\begin{align}\label{Equ:Relative3}
r(0) =\sum_{i = 0}^{\bar{\tau}} \alpha_i^2 p_i(1-p_i).
\end{align}

According to Lemma \ref{Lma:omega_i}.3 and (\ref{Equ:Relative}), letting $k_1=k$,
$k_2=k+l$, $i_1=k-n$ and $i_2=k+l-n$ yields
\begin{align}\label{Equ:Relative2}
r(l) =-\sum_{i = 0}^{\bar{\tau}-l} \alpha_i\,\alpha_{i+l}\,p_i\,p_{i+l}, \;\; 0<l \leq \bar{\tau}.
\end{align}
It holds for  $l > \bar{\tau}$ that $\omega(k,n)\omega(k+l,n) \equiv 0$.
Hence, $r(l) \equiv 0$ for $l > \bar{\tau}$.

In the case when $l<0$, note the fact that $ \omega(k+l,n) \equiv 0  $ for any $k+l<n$. It is verified by (\ref{Equ:Relative}) that $r(l)=r(-l).$}

Subsequently, define the energy spectral density of the channel uncertainty $\Omega$ as follows:
\begin{align}\label{Equ:Spectral_Density}
S_\Omega(z) = \sum_{l=-\infty}^{\infty}r(l)z^{-l}.
\end{align}

\begin{lem}\label{lemma2}
The energy spectral density $S_\Omega(z)$ of the channel uncertainty $\Omega$ can be written as
\begin{equation}\label{Equ:PSD_Omega}
\begin{aligned}
S_\Omega=\frac{1}{2} \sum_{i_1,i_2=0}^{\bar{\tau}}(\alpha_{i_1}z^{i_1}-\alpha_{i_2}z^{i_2})
(\alpha_{i_1}z^{-i_1}-\alpha_{i_2}z^{-i_2})p_{i_1}p_{i_2}.
\end{aligned}
\end{equation}
\end{lem}
\begin{proof}
{Note the fact that any $|l|>\bar{\tau}$, $r(l)\equiv 0$. It holds that
\begin{align*}
S_\Omega(z) = \sum_{l=-\bar{\tau}}^{\bar{\tau}}r(l)z^{-l}.
\end{align*}
}

It follows from the definition (\ref{Equ:Spectral_Density}) of $S_\Omega(z)$, $r(0)$ and $r(l)$ given in
(\ref{Equ:Relative3}) and (\ref{Equ:Relative2}), respectively, that
\begin{equation}\label{Equ:X2}
\begin{aligned}
S_\Omega(z)=&\sum_{i=0}^{\bar{\tau}} \alpha_i^2 p_i(1-p_i)
- \sum_{\begin{subarray}{c} i_1,i_2=0\\ i_1\neq i_2  \end{subarray}  }^{\bar{\tau}}  \alpha_{i_1}\alpha_{i_2}
p_{i_1}p_{i_2}z^{i_1-i_2}.
\end{aligned}
\end{equation}
Note the fact that ${\displaystyle \sum_{i=0}^{\bar{\tau}} p_{i}=1}$. It holds that
\begin{equation}\label{Equ:X3}
\begin{aligned}
\sum_{i=0}^{\bar{\tau}} \alpha_i^2 p_i(1-p_i)&=\sum_{i_1=0}^{\bar{\tau}} \alpha_{i_1}^2
p_{i_1}\sum_{i_2=0}^{\bar{\tau}} p_{i_2} - \sum_{i=0}^{\bar{\tau}} \alpha_i^2 p_i^2\\
&\hspace{0cm}=\frac{1}{2}\sum_{i_1=0}^{\bar{\tau}}\sum_{i_2=0}^{\bar{\tau}}(\alpha_{i_1}^2+\alpha_{i_2}^2)
p_{i_1} p_{i_2} - \sum_{i=0}^{\bar{\tau}} \alpha_i^2 p_i^2
\end{aligned}
\end{equation}
and
\begin{equation}\label{Equ:X4}
\begin{aligned}
&\sum_{\begin{subarray}{c} i_1,i_2=0\\ i_1\neq i_2  \end{subarray}  }^{\bar{\tau}}  \alpha_{i_1}\alpha_{i_2}
p_{i_1}p_{i_2}z^{i_1-i_2}
=\sum_{i_1,i_2=0}^{\bar{\tau}} \alpha_{i_1}\alpha_{i_2} p_{i_1}p_{i_2}z^{i_1-i_2} - \sum_{i=0}^{\bar{\tau}}
\alpha_i^2 p_i^2.
\end{aligned}
\end{equation}
Substituting \eqref{Equ:X3} and \eqref{Equ:X4} into \eqref{Equ:X2} leads to \eqref{Equ:PSD_Omega}.
\end{proof}

\hlt{
\begin{remark}
Note, from \eqref{Equ:Relative3}-\eqref{Equ:Relative2}, that the autocorrelation $r(l)$ of the subsequence $\{\omega(k,n): -\infty < k < \infty\}$ is only dependent on the parameters of the transmission block (i.e., the PMF of the delay and the given weights in the receiving strategy) but independent of $n$, so is the corresponding energy spectral density function $S_\Omega$ given by \eqref{Equ:Spectral_Density}.
This allows us to establish the input-output relation of the channel uncertainty $\Omega$.
\end{remark}
}


{
\begin{lem} \label{Lem:u_independent_tau}
Suppose that Assumptions \ref{Assump:Sum_of_probability} and \ref{Assump:independent_assumption} hold for the channel induced delay process and the external input sequence of the system.
Then for any $n_1 \ge n_2 \ge 0$, $\tau_{n_1}$ is independent of the channel input $u(n_1)$ and $u(n_2)$ in the system.
\end{lem}

\begin{proof}
Since the plant $P$ is assumed to be strictly proper, the controller output $u(n)$ only depends on the past inputs of $P$, which is determined by $\{v(0),\cdots,v(n-1)\}$ and $\{\tau_{0},\cdots,\tau_{n-1}\}$, provided that $P$ and $K$ are relaxed at $n=0$.
Then by Assumptions \ref{Assump:Sum_of_probability} and \ref{Assump:independent_assumption}, the current channel transmission time $\tau_n$ is independent of the current and past channel inputs, which completes the proof.
\end{proof}
}

Define the autocorrelation of the sequence $\{d(0), d(1), \cdots\}$ by
\begin{align}\label{d_auto_correlation}
r_d(l)=\left\{
\begin{matrix}
{\displaystyle \lim_{\bar{k}\rightarrow \infty}\frac{1}{\bar{k}+1} \sum_{k=0}^{\bar{k}}\E\{d(k)d(k+l)\}},
& 0 \leq l < \infty \\
{\displaystyle \lim_{\bar{k}\rightarrow \infty}\frac{1}{\bar{k}+1} \sum_{k=-l}^{\bar{k}-l}\E\{d(k)d(k+l)\}},
& -\infty <l<0
\end{matrix}
\right.
\end{align}
It follows from (\ref{d_auto_correlation}) that
\begin{align}
{r_d(l)=r_d(-l), \; l=-1,-2, \cdots.}
\end{align}
The power spectral density of $d(k)$ is as follows:
\begin{align}\label{S_d_definition}
S_d(z)=\sum_{l=-\infty}^{\infty}r_d(l)z^{-l}.
\end{align}

Denote the $i$th-component of $d(k)$ in \eqref{Equ:d_total} by $d_i(k)$, $i \in {\mathcal D}$, i.e.,
$$d_i(k)=\omega(k, k-i)u(k-i), \quad     {i \in \mathcal{D}}.$$

\begin{lem}\label{Lma:correlation_di_dj}
Suppose that Assumptions \ref{Assump:Sum_of_probability} and \ref{Assump:independent_assumption} hold
for the channel induced delay process {and the external input sequence of the \hltb{system}}.
It holds for $k_1$, $k_2=0,1,2,\cdots$ that
\begin{enumerate}
\item for $i \in \mathcal{D}$,
\begin{align}\label{Equ:Expectation_of_dm_dm}
\E\left\{ d_i(k_1) d_i(k_2) \right\}
=\delta (k_1 - k_2) \alpha_i^2 p_i{(1 - p_i)} \E\left\{ u^2( k_1-i ) \right\};
\end{align}
\item   for $i_1 \ne i_2$, $i_1, i_2 \in \mathcal{D}$,
\begin{align}\label{Equ:Expectation_of_dm_dn}
&\E\left\{ d_{i_1}(k_1) d_{i_2}(k_2) \right\} \\
&\hspace{1cm}= -\delta(k_1-i_1-k_2+i_2) \alpha_{i_1} \alpha_{i_2} p_{i_1} p_{i_2} \E\left\{ u^2(k_1-i_1) \right\}.\nonumber
\end{align}
\end{enumerate}
\end{lem}
\begin{proof}
From Assumption \ref{Assump:Sum_of_probability} and \ref{Assump:independent_assumption},
for any $k_1\neq k_2 \geq 0$ and any $i\geq 0$, $\omega(k_1, k_1-i)$ is independent of
$\omega(k_2,k_2-i)$ and $u(k_1-i)$. Without loss of generality, assume $k_1 > k_2$.
Then $\omega(k_1, k_1-i)$ is independent of $u(k_2-i)$. This leads to
\begin{align*}
&\E\left\{ d_i(k_1) d_i(k_2) \right\}\nonumber\\
=& \E\left\{\omega(k_1, k_1-i)\right\}\E\left\{ u(k_1-i) \omega(k_2, k_2-i)u(k_2-i) \right\}\\
=&0.
\end{align*}
On the other hand, it holds that
\begin{align*}
\E\left\{ d_i(k_1) d_i(k_1) \right\}
= \E\left\{\omega^2(k_1, k_1-i)\right\}\E\left\{ u^2(k_1-i) \right\}.
\end{align*}
Consequently, from Lemma \ref{Lma:omega_i}.2, Lemma \ref{Lma:correlation_di_dj}.1 holds.

Now, we prove Lemma \ref{Lma:correlation_di_dj}.2. Note that $\omega(k_1, k_1-i_1)$ is independent of $\omega(k_2,
k_2-i_2)$ for $k_1-i_1 \ne k_2-i_2$. So, it holds that
\[
\E\left\{ d_{i_1}(k_1) d_{i_2}(k_2) \right\}=0.
\]
For the case $k_1-i_1 = k_2-i_2$, $\omega(k_1,k_1-i_1)$ and $\omega(k_2, k_2-i_2)$ are independent of
$u(k_1-i_1)$, so we have that
\begin{align}\label{d_d_correlation_12}
\E\left\{ d_{i_1}(k_1) d_{i_2}(k_2) \right\}
=\E\left\{\omega(k_1, k_1-i_1)\omega(k_2, k_1-i_1)\right\}\E\left\{ u^2(k_1-i_1) \right\}.
\end{align}
Applying Lemma \ref{Lma:omega_i}.3 to (\ref{d_d_correlation_12}) leads to (\ref{Equ:Expectation_of_dm_dn}).
Proof is completed.
\end{proof}

For any stochastic sequence $\{u(0), u(1), u(2), \cdots\}$, denote its averaged power by $\|u\|_{\cal P}$,
\begin{align*}
\|u\|_{{\cal P}}^2=\lim_{\bar{k}\rightarrow \infty}\frac{1}{\bar{k}+1} \sum_{k=0}^{\bar{k}}\E\{u^2(k)\}.
\end{align*}

\begin{lem}\label{d_PSD}
Suppose that Assumptions \ref{Assump:Sum_of_probability} and \ref{Assump:independent_assumption} hold
for the channel induced delay process and {the external input sequence of the system}.
The power spectral density $S_d(z)$ of the channel uncertainty $\Omega$'s
output $\{d(0), d(1), d(2), \cdots\}$ is given by
\begin{align}\label{S_d_S_omega}
S_d(z)=S_{\Omega}(z) \|u\|_{\cal P}^2.
\end{align}
\end{lem}

\begin{proof}
The autocorrelation of $d(k)$ is determined by the autocorrelations of its components $d_i(k)$.
It follows from \eqref{Equ:d_total} that
\begin{align}\label{d_d_correlation}
\hspace{-.1cm}\E\{d^2(k)\}&=\sum_{i_1=0}^{\bar{\tau}}\sum_{i_2=0}^{\bar{\tau}}\E\{d_{i_1}(k)d_{i_2}(k)\}\\
\hspace{-.1cm}&=\sum_{i_1=0}^{\bar{\tau}}\E\{d_{i_1}^2(k)\}+\sum_{\begin{subarray}{c} i_1,i_2=0 \\ i_1\neq
i_2\end{subarray}}^{\bar{\tau}}\E\{d_{i_1}(k)d_{i_2}(k)\}.\nonumber
\end{align}
Applying Lemma \ref{Lma:correlation_di_dj} to (\ref{d_d_correlation}) leads to
\begin{align}\label{d_d_correlation_1}
\E\{d^2(k)\}&=\sum_{i_1=0}^{\bar{\tau}}\E\{d_{i_1}^2(k)\}\\
&=\sum_{i=0}^l \alpha_i^2 p_i(1-p_i) \E\{u^2(k-i)\}.\nonumber
\end{align}

For $1 \leq l \leq {\bar{\tau}}$, it holds that
\begin{align}\label{d_d_correlation2}
\E\{d(k)d(k+l)\}
=&\sum_{i_1=0}^{\bar{\tau}}\sum_{i_2=0}^{\bar{\tau}}\E\{d_{i_1}(k)d_{i_2}(k+l)\}\\
=&\sum_{i=0}^{\bar{\tau}}\E\{d_i(k)d_i(k+l)\}+\sum_{\begin{subarray}{c} i_1,i_2=0 \\ i_1\neq i_2 \\ i_2\neq i_1+l
\end{subarray}}^{\bar{\tau}}\E\{d_{i_1}(k)d_{i_2}(k+l)\}\nonumber\\
&+\sum_{\begin{subarray}{c} i_1,i_2=0 \\ i_1=i_2+l\end{subarray}}^{\bar{\tau}}\E\{d_{i_1}(k)d_{i_2}(k+l)\}\nonumber
\end{align}
According to Lemma \ref{Lma:correlation_di_dj}, we write (\ref{d_d_correlation2}) as
\begin{align}\label{d_d_correlation3}
\E\{d(k)d(k+l)\}
=-\sum_{i=0}^{{\bar{\tau}}-l} {\alpha_i\, \alpha_{i+l}\, p_i\, p_{i+l}} \E\left\{ u^2(k-i) \right\}.
\end{align}
Moreover, for any $l>{\bar{\tau}}$ and $i_1, i_2\in {\mathcal D}$, $\omega(k+l,k+l-i_2)$ is independent of
$\omega(k,k-i_1)$. It leads to
\begin{align}\label{d_d_correlation4}
\E\{d(k)d(k+l)\}=0.
\end{align}

Substituting (\ref{d_d_correlation_1}), (\ref{d_d_correlation3}) and (\ref{d_d_correlation4}) into
(\ref{d_auto_correlation}), respectively, we obtain that
\begin{align}\label{d_d_correlation5}
r_d(0)=\sum_{i=0}^{\bar{\tau}} \alpha_i^2 p_i(1-p_i) \lim_{\bar{k}\rightarrow
\infty}\frac{1}{\bar{k}+1}\sum_{k=0}^{\bar{k}}\E\{u^2(k-i)\},
\end{align}
\begin{align}\label{d_d_correlation6}
r_d(l)&=-\sum_{i=0}^{\bar{\tau}} {\alpha_i \,\alpha_{i+l}\, p_i \,p_{i+l}} \lim_{\bar{k}\rightarrow
\infty}\frac{1}{\bar{k}+1}\sum_{k=0}^{\bar{k}}\E\{u^2(k-i)\}, \;
1\leq l \leq  {\bar{\tau}},
\end{align}
and
\begin{align}\label{d_d_correlation9}
r_d(l)=0, \; l > \bar{\tau}.
\end{align}

For any bounded sequence $\{ \E\{u^2(k-i)\},\; k=0,1,2,\cdots, \}$, $i\in {\mathcal D}$, it holds that
\begin{align}\label{u_power}
\lim_{\bar{k}\rightarrow \infty}\frac{1}{\bar{k}+1}\sum_{k=0}^{\bar{k}}\E\{u^2(k-i)\}=\|u\|_{\cal P}^2.
\end{align}
Taking account to (\ref{u_power}), (\ref{Equ:Relative3}) and (\ref{Equ:Relative2}), we rewrite
(\ref{d_d_correlation5}) and (\ref{d_d_correlation6}) as follows:
\begin{align}\label{d_d_correlation7}
r_d(0)=\|u\|_{\cal P}^2\sum_{i=0}^{\bar{\tau}} \alpha_i^2 p_i(1-p_i)
=\|u\|_{\cal P}^2 r(0),
\end{align}
and
\begin{align}\label{d_d_correlation8}
r_d(l)=-\|u\|_{\cal P}^2\sum_{i=0}^{\bar{\tau}} {\alpha_i \,\alpha_{i+l}\, p_i\, p_{i+l}}
=\|u\|_{\cal P}^2 r(l), \; 1\leq l \leq  {\bar{\tau}},
\end{align}
respectively.

Substituting (\ref{d_d_correlation7}), (\ref{d_d_correlation8}) and (\ref{d_d_correlation9}) into
(\ref{S_d_definition}) results in
\begin{align}\label{S_d_definition_A}
S_d(z)=\|u\|_{\cal P}^2 \sum_{l=-{\bar{\tau}}}^{\bar{\tau}}r(l)z^{-l}.
\end{align}
Applying (\ref{Equ:Spectral_Density}) into (\ref{S_d_definition_A}) leads to (\ref{S_d_S_omega}).
\end{proof}

\begin{remark}
As mentioned in the preceding section, an unreliable channel with random packet dropout can be modeled
by a multiplicative white noise process with variance \hlt{$\sigma^2$}, and the variance \hlt{$\sigma^2$} is determined by
the packet dropout probability (see \cite{Elia2005Remote, Elia2011limitations, Gu2016An, Su2016Control}
and references therein). According to our framework, it holds for this case that $\bar{\tau}=1$, the mean
channel $H$ is a constant determined by the packet dropout probability and the spectral density
$S_{\Omega}$ of the channel uncertainty is $\sigma^2$. For this white noise channel uncertainty, it holds
that $S_d(z)=\|d\|_{\cal P}^2$. It follows from Lemma \ref{d_PSD} that
\begin{align}\label{frequency_input_output_MN}
\|d\|_{\cal P}^2=\sigma^2\|u\|_{\cal P}^2.
\end{align}
Moreover, it is not hard to verify that (\ref{frequency_input_output_MN}) holds for all channel uncertainties
modeled by multiplicative white noises with zero mean and variance $\sigma^2$\emph{}.
\end{remark}

It is well-known that for spectral density $S_{\Omega}$, there exists a minimum phase polynomial $\Phi(z)$
of $z^{-1}$ with degree $\bar{\tau}$ and real coefficients satisfying
\begin{align}\label{Equ:Spectral_factization}
S_\Omega(z)=\Phi(z^{-1})\Phi(z).
\end{align}
It is referred to as the spectral factorization of $S_\Omega(z)$ in literatures (see for example \cite{Francis1987, Zhou1995}).

Notice the fact that $\Phi(z)$ and $H(z)$ are real polynomials of $z^{-1}$ with degree $\bar{\tau}$. The function
\begin{align}\label{Equ:Deviation-function}
W(z) := {\Phi(z)}{H^{-1}(z)}
\end{align}
is, therefore, proper and real-rational. The complementary sensitivity function in the nominal system $G$ shown in
Fig. \ref{Fig:System_G_with_uncertainty} is given by
\begin{align}\label{Equ:Sensitivity}
T(z)={H(z)K(z)P(z)}{[1-H(z)K(z)P(z)]}^{-1}.
\end{align}
The next theorem establishes a new small gain theorem for \msis of the networked feedback system shown in Fig. \ref{Fig:System_connected_channel}.

\begin{thm}\label{Thm:Mean_square_IO_stable}
Under Assumptions \ref{Assump:Sum_of_probability}, \ref{Assump:independent_assumption}, and \ref{Assump:K_stabilizes_G}, the networked feedback system \hlt{with a given linear receiving strategy} and a controller $K \in {\cal K}$ is mean-square \hlt{input-output} stable if and only if
\begin{equation}\label{Equ:Mean-square_IO_stable}
\|W(z)T(z)\|_2^2 < 1.
\end{equation}
\end{thm}

\begin{proof}
Consider the system in Fig. \ref{Fig:G_Omega}. Denote the power spectral densities of the signals $u$, $e$,
$d$ and $v$ by $S_u(z)$, $S_e(z)$, $S_d(z)$ and $S_v(z)$, respectively. Denote the autocorrelation of
$u$ by $r_u(l), l=0,\pm 1,\pm 2,\cdots$.  It holds for the averaged power and power spectral density of the signal
$u$ that
\begin{align}\label{u_power_auto_correleation}
\|u\|_{\cal P}^2=r_u(0)
\end{align}
and
\begin{align}\label{auto_spectral}
r_u(0)=\frac{1}{2\pi}\int_{-\pi}^{\pi} S_u(e^{j\theta})d\theta.
\end{align}
It is well-known that for the linear system $G$, the power spectral densities of its input signal $e$ and output
signal $u$ satisfy
\begin{align}\label{PSD_linear system}
S_u(e^{j\theta})=G(e^{j\theta})S_e(e^{j\theta})G^*(e^{j\theta}).
\end{align}
Hence, we obtain that
\begin{align}\label{u_power_spectral}
\|u\|_{\cal P}^2=\frac{1}{2\pi}\int_{-\pi}^{\pi} G(e^{j\theta})S_e(e^{j\theta})G^*(e^{j\theta})d\theta.
\end{align}
{Since the input sequence $\{v(n), n=0,1,2, \cdots \}$ is independent of $\{\tau_n, n=0,1,2, \cdots \}$,
for any $k_1$, $k_2$ and $i$, $v(k_1)$ and $\omega(k_2, k_2-i)$ are mutually \hltb{independent},
$\omega(k_2,k_2-i)$ is independent of $u(k_2-i)$}. This leads to that
\begin{align}
\E\left\{ v(k_1)d(k_2) \right\}
=\sum_{i=0}^{\bar{\tau}}\E\left\{v(k_1)\omega(k_2,k_2-i)u(k_2-i)\right\}=0.
\end{align}
Consequently, we obtain that
\begin{align}\label{e_PSD}
S_e(e^{j\theta})=S_v(e^{j\theta})+S_d(e^{j\theta}).
\end{align}
Applying Lemma \ref{d_PSD}, we write (\ref{e_PSD}) as
\begin{align}\label{e_PSD_uP}
S_e(e^{j\theta})=S_v(e^{j\theta})+S_{\Omega}(e^{j\theta})\|u\|_{\cal P}^2.
\end{align}
Substituting (\ref{e_PSD_uP}) into (\ref{u_power_spectral}) leads to
\begin{align}\label{u_power_spectral_A}
\|u\|_{\cal P}^2=&\frac{1}{2\pi}\int_{-\pi}^{\pi}
G(e^{j\theta})S_{\Omega}(e^{j\theta})G^*(e^{j\theta})d\theta\|u\|_{\cal P}^2\\
&\hspace{2cm}+\frac{1}{2\pi}\int_{-\pi}^{\pi} G(e^{j\theta})S_v(e^{j\theta})G^*(e^{j\theta})d\theta.\nonumber
\end{align}
Taking account to the spectral factorization (\ref{Equ:Spectral_factization}), we have that
\begin{align}\label{u_power_spectral_B}
\|u\|_{\cal P}^2=& \|G({z})\Phi({z})\|_2^2\|u\|_{\cal P}^2
\hspace{0cm}+\frac{1}{2\pi}\int_{-\pi}^{\pi} G(e^{j\theta})S_v(e^{j\theta})G^*(e^{j\theta})d\theta.
\end{align}
The power $\|u\|_{\cal P}$ exists if and only if $\|G({z})\Phi({z})\|_2^2<1$.

Note that
\[
G({z})\Phi({z})=W({z})T({z}).
\]
This theorem holds.
\end{proof}

\begin{remark}
In literatures (see for example \cite{Elia2005Remote} and references therein), there is a classical version of the mean-square small gain theorem for a networked feedback system over unreliable channel of which the
channel uncertainty is modeled as a multiplicative white noise process. {As shown by (\ref{Equ:Classical_mean-square-stability}), the mean-square (input-output) stability of the system is determined by the interaction between the {\em  coefficient of variation} of the unreliable channel's gain and the complementary sensitivity function of the nominal system.
Theorem \ref{Thm:Mean_square_IO_stable} is a generalized version of this \msis criterion for the networked system with a random channel induced \hltb{delay}.
As shown by the inequality (\ref{Equ:Mean-square_IO_stable}), the \msis of the \hltb{networked system} is determined by the interaction between the factor $W$ and the complementary sensitivity function of the nominal system.
In fact, $W(e^{j\theta}) = \Phi(e^{j\theta})/H(e^{j\theta})$ is the {\em coefficient of variation} of the channel gain at the given frequency $\theta$.
Here, $W$ is referred to as the {\em frequency response of variation} of the channel.

To understand the role \hltb{played by} $W$ in data transmission, we consider the channel with a random channel induced delay shown in Fig. \ref{Fig:System_G_with_uncertainty}
where $H$ and $\Omega$ are the mean channel and the channel uncertainty, respectively. Since the mean channel $H$ is a linear time-invariant system,
the power spectral density $S_{\bar{u}}$ of its output $\bar{u}$ \hltb{satisfies}
that
\begin{align*}
S_{\bar{u}}(e^{j\theta})=H(e^{j\theta})S_u(e^{j\theta})H^*(e^{j\theta}).
\end{align*}
The power spectral density $S_d(e^{j\theta})$ of the output of $\Omega$ is given by Lemma \ref{d_PSD}.
Thus, we obtain that
\begin{align}\label{SNR}
\frac{S_{\bar{u}}(e^{j\theta})}{S_d(e^{j\theta})}
=\frac{H(e^{j\theta})H^*(e^{j\theta})}{S_{\Omega}(e^{j\theta})}
\frac{S_{u}(e^{j\theta})}{\|u\|_{\cal P}^2}=\frac{1}{W(e^{j\theta})W^*(e^{j\theta})}\frac{S_{u}(e^{j\theta})}{\|u\|_{\cal P}^2}.
\end{align}
Note from the \hltb{structure} of the channel (or (\ref{Equ:Channel_output})) that the ratio ${\displaystyle \frac{S_{\bar{u}}(e^{j\theta})}{S_d(e^{j\theta})}}$ is
the signal-to-noise ratio (SNR) of the channel at the frequency $\theta$ and
${\displaystyle \frac{S_{u}(e^{j\theta})}{\|u\|_{\cal P}^2}}$ is
the normalized power spectral density of the channel input $u$.
From (\ref{SNR}), we can see that
increasing the power $\|u\|_{\cal P}$ of the input signal could not yield a
\hltb{greater} SNR. For a channel input with a given normalized power spectral density,
the SNR of the channel and the frequency response of variation of the channel
are inversely proportional at any given frequency $\theta$.
}
\end{remark}

{
\begin{remark}
In \cite{MassSilva2014}, it is shown that for the mean-square (input-output) stability of a networked feedback system, there is an equivalence between a fading channel and an \hltb{additive white} noise channel with a signal-to-noise ratio \hltb{constraint}. Comparing Theorem \ref{Thm:Mean_square_IO_stable} with the inequality (15) of Theorem 1 in \cite{Franceschetti2012} and the mean-square stabilizability condition (3) in \cite{GonzalezChen2019}, we can see that there is a similar \hltb{equivalence} between \hltb{a} channel with random data transmission delays  and \hltb{an additive colored} noise channel with a signal-to-noise ratio constraint. More precisely, for the \msis of the networked feedback system, the channel uncertainty of which the frequency response of variation is given by $W$ is equivalent to \hltb{an additive colored} noise in an additive noise channel with a signal-to-noise ratio constraint, where the power spectral density of the noise is given by $W^*(e^{j\theta})W(\theta^{j\theta})$ and the upper bound of the signal-to-noise ratio of the channel is given by $\|W\|_2^{-2}$.
\end{remark}
}


\section{{{Frequency Response of Variation}} vs. Unstable Poles in the Mean-square Input-output Stabilizability}\label{FSNR_UP}

In this part, a criterion of the mean-square input-output stabilizability via output feedback is studied for the networked feedback system. We attempt to precisely explain the inherent connection between the mean-square input-output stabilizability of the system, the {{frequency response of variation $W$}} of the unreliable channel and the unstable poles of the plant $P$. To seek a simplicity, it is assumed that the plant is minimum-phase and with a relative degree $\tau
> 0$.

{
From the stability criterion \eqref{Equ:Mean-square_IO_stable}, the mean-square input-output stabilizability condition of the
networked feedback system is straightforwardly obtained.
\begin{lem}\label{Lma:Stabilizability_condition}
Under Assumptions \ref{Assump:Sum_of_probability}, \ref{Assump:independent_assumption}, and \ref{Assump:K_stabilizes_G}, the networked feedback system \hlt{with a random channel induced delay and a given linear receiving strategy, as shown in Fig. \ref{Fig:System_connected_channel},}
is mean-square input-output stabilizable if and only if
\begin{equation}\label{Equ:Stabilizable_condition_1}
\inf_{K \in \mathcal{K}}\|W(z)T(z)\|_2^2 < 1.
\end{equation}
\end{lem}
\begin{proof}
See \cite[Lemma 2]{Qi2017Control}.
\end{proof}
}

Let a coprime factorization of the SISO plant transfer function $H(z)P(z)$ be given by
\begin{equation*}
H(z)P(z) = N M^{-1},
\end{equation*}
where $N,M \in \mathcal{RH}_\infty$ satisfy the B\'{e}zout's identity
\begin{equation}\label{Equ:Bezout_indentity}
MX + NY = 1,
\end{equation}
for some $X,Y \in \mathcal{RH}_\infty$. It is well-known that the set of all stabilizing feedback controllers to
$H(z)P(z)$ is parameterized as (see \cite{Francis1987, Zhou1995})
\begin{equation}\label{Equ:Youla_param_controller}
{\cal K} = \left\{\left.K=-{(Y +MQ)}{(X- NQ)^{-1}}\right| Q \in \mathcal{RH}_\infty \right\}.
\end{equation}
Applying a stabilizing controller from the set ${\cal K}$ to the networked feedback system, we have that
\begin{equation}\label{Equ:WT_Q}
{W(z){T}\left( z \right)} = - W(Y + M Q)N.
\end{equation}
In light of Lemma \ref{Lma:Stabilizability_condition},
the following condition for the mean-square input-output stabilizability of the system is immediate.

\begin{lem}\label{Lma:Stabilizable_condition}
Under Assumptions \ref{Assump:Sum_of_probability}, \ref{Assump:independent_assumption}, and \ref{Assump:K_stabilizes_G}, the networked feedback system \hlt{with a random channel induced delay and a given linear receiving strategy, as shown in Fig. \ref{Fig:System_connected_channel},}
is mean-square input-output stabilizable if and only if
\begin{equation}\label{Equ:Stabilizable_condition}
\inf\limits_{Q \in \mathcal{RH}_\infty}\left\| W(Y + M Q)N \right\|_2^2 < 1.
\end{equation}
\end{lem}

{
\begin{proof}
Substituting \eqref{Equ:WT_Q} into Lemma \ref{Lma:Stabilizability_condition} completes the proof.
\end{proof}
}

As we can see in Lemma \ref{Lma:Stabilizable_condition}, the solution to the minimization problem in
\eqref{Equ:Stabilizable_condition} requires synthesizing an optimal $ Q \in \mathcal{RH}_\infty $.
To this end, an inner-outer factorization of $M(z)$ is considered. Suppose that $\lambda_1$, $\cdots$, $\lambda_n\in
\bar{\mathbb{D}}^c$ are all unstable poles of $P(z)$, i,e, these are zeros of $M(z)$. An inner-outer factorization
of $M(z)$ is given by
\begin{align}\label{inner_outer}
M(z) = M_{in}(z) M_{out}(z)
\end{align}
where
\begin{equation}\label{Equ:Factorize_M}
\begin{aligned}
M_{in} &= \prod_{i=1}^{n}M_{in,i},~~~M_{in,i}=\frac{z - \lambda_i}{1-{\lambda}_i^* z}.
\end{aligned}
\end{equation}
For any scalar real parameter inner $M_{in}(z)$,
let $M_{in}^{\sim}=M_{in}(z^{-1})$. It holds that
$M_{in}^{\sim}M_{in}=1$.
\begin{lem}\label{Balanced_realization_Lemma}
For the inner $M_{in}$ given in (\ref{Equ:Factorize_M}), there exists a balanced realization of
$M_{in}=\sys{A_{in}}{B_{in}}{C_{in}}{D_{in}}$ such that
\begin{align}\label{Balanced_realization}
\begin{bmatrix}
A_{in}  &  B_{in}  \\
C_{in}  &  D_{in}
\end{bmatrix}
\begin{bmatrix}
A_{in}^*  &  C_{in}^*  \\
B_{in}^*  &  D_{in}^*
\end{bmatrix}
=
\begin{bmatrix}
I  &  0  \\
0  &  1
\end{bmatrix}
.
\end{align}
\end{lem}

{
\begin{proof}
See \cite[Corollary 21.16]{Zhou1995}.
\end{proof}
}

\begin{lem}\label{inverse_realization}
For any LTI system $G=\sys{A}{B}{C}{D}$ with invertible $D$, its inverse is given by
\begin{align}\label{inverse_system}
G^{-1}=\sys{A-BD^{-1}C}{-BD^{-1}}{D^{-1}C}{D^{-1}}.
\end{align}
\end{lem}

{
\begin{proof}
See \cite[Lemma 3.15]{Zhou1995}.
\end{proof}
}

Applying Lemmas \ref{Balanced_realization_Lemma} and \ref{inverse_realization} results in
\begin{align}\label{M_in_inv_realization}
M_{in}^{-1}=\sys{A_{in}^{-*}}{-B_{in}D_{in}^{-1}}{D_{in}^{-1}C_{in}}{D_{in}^{-1}}.
\end{align}

Now, we are ready to present the mean-square input-output stabilizability criterion for the networked feedback system in terms
of the interaction between unstable poles $\lambda_1, \cdots, \lambda_n$ of the plant and the {{frequency response of variation $W$}}.

\begin{thm}\label{Thm:Mean_squre_bounds}
Suppose that the plant $P(z)$ of the networked feedback system \hlt{with
a random channel induced delay and a given linear receiving strategy} shown
in Fig. \ref{Fig:System_connected_channel} is with a
relative degree $\tau > 0$ and satisfies Assumption \ref{Assump:K_stabilizes_G}, the unreliable channel with a random channel delay satisfies Assumptions \ref{Assump:Sum_of_probability} and \ref{Assump:independent_assumption}. Let $\lambda_i \in \bar{\mathbb{D}}^c,i=1,\cdots,n$ be unstable poles
of $P(z)$ and  $M_{in}=\sys{A_{in}}{B_{in}}{C_{in}}{D_{in}}$ be the associated balanced realization. Then the
networked system is mean-square input-output stabilizable if and only if
\begin{equation}\label{Equ:Existence_condition}
D_{in}^{-1}C_{in}{A_{in}^{-*}}^{\tau-1}W(A_{in}^{-*}) W(A_{in}^{-1}) A_{in}^{-(\tau-1)} C_{in}^* D_{in}^{-*} < 1.
\end{equation}
\end{thm}

\begin{proof}
It follows from Lemma \ref{Lma:Stabilizable_condition} that the key in the proof is to find the optimal solution
to the minimization problem ${\displaystyle \inf_{Q \in \mathcal{RH}_\infty}\left\| W(Y + M Q)N \right\|_2^2}$.

With this purpose, we apply the inter-outer factorization (\ref{inner_outer}) into (\ref{Equ:Bezout_indentity})
and obtain that
\begin{align}\label{outerX}
M_{out}X=M_{in}^{-1}(1-NY).
\end{align}
Let the impulse responses of the functions $M_{in}^{-1}$ and $M_{out}X$ be $\{\hat{m}_{in}(0), \hat{m}_{in}(1),
\cdots\}$ and $ \{\hat{m}_{outX}(0),\hat{m}_{outX}(1),\cdots\}$, respectively, i.e.,
\begin{align*}
M_{in}^{-1}&=\hat{m}_{in}(0)+\hat{m}_{in}(1)z^{-1}+\hat{m}_{in}(2)z^{-2}+\cdots \\
M_{out}X&=\hat{m}_{outX}(0)+\hat{m}_{outX}(1)z^{-1}+\hat{m}_{outX}(2)z^{-2}\!+\!\cdots .
\end{align*}
Note the fact that the relative degrees of $M_{in}^{-1}$ and $M_{out}X$ are zero and the relative degree of $N$ is
$\tau$. It follows from (\ref{outerX}) that
\begin{align}\label{M_in_tau_impulse}
\hat{m}_{in}(k)=\hat{m}_{outX}(k), \; k=0,1,\cdots \tau-1.
\end{align}
Let
\begin{align}\label{M_in_tau}
\hat{M}_{in,\tau}=\hat{m}_{in}(0)+\hat{m}_{in}(1)z^{-1}+\cdots+\hat{m}_{in}(\tau-1)z^{-\tau+1}.
\end{align}

Applying (\ref{Equ:Bezout_indentity}) and identity $z^{-\tau}z^{\tau}=1$, we have that
\begin{align}\label{Model_match2}
&\| {W(Y +  MQ)N}\|_2^2
= \left\| {W(1-MX + MQN){z^\tau }} \right\|_2^2.
\end{align}
Due to the identity $M_{in}^{\sim}M_{in}=1$, it holds that
\begin{align}\label{Model_match3}
&\left\| W(1-MX + MQN){z^\tau } \right\|_2^2\\
&\hspace{1cm}=\left\| W(M_{in}^{-1}-M_{out}X + M_{out}QN){z^\tau } \right\|_2^2\nonumber\\
&\hspace{1cm}=\left\| W(M_{in}^{-1}-\hat{M}_{in,\tau}){z^\tau }+W(\hat{M}_{in,\tau}-M_{out}X + M_{out}QN) z^\tau  \right\|_2^2.\nonumber
\end{align}

From (\ref{M_in_tau_impulse}), one can see that $\hat{M}_{in,\tau}-M_{out}X$ is with relative degree $\tau$ and
\begin{align}
(\hat{M}_{in,\tau}-M_{out}X)z^{\tau} \in \mathcal{RH}_2.
\end{align}

On the other hand, $W(M_{in}^{-1}-\hat{M}_{in,\tau}){z^\tau }$ can be decomposed as a summation of two functions
$Z_1$, $Z_2$ from $\mathcal{RH}_2$ and $\mathcal{RH}_2^{\perp}$, respectively, i.e.
\begin{align}\label{WM_decompose}
W(M_{in}^{-1}-\hat{M}_{in,\tau}){z^\tau }=Z_1+Z_2,\; Z_1\in \mathcal{RH}_2, \; Z_2 \in \mathcal{RH}_2^{\perp}.
\end{align}
Hence, (\ref{Model_match3}) is written as
\begin{align}\label{Model_match4}
&\left\| W(1-MX + MQN){z^\tau } \right\|_2^2\\
=&\left\| Z_2\right\|_2^2+\left\|Z_1+W(\hat{M}_{in,\tau}-M_{out}X + M_{out}QN) z^\tau  \right\|_2^2.\nonumber
\end{align}

Since $N$ has no non-minimum phase zeros and relative degree $\tau$,
selecting a proper $Q\in \mathcal{RH}_{\infty}$ leads to
\begin{equation}\label{Equ:Optimal_Q}
\left\| Z_1+W(\hat{M}_{in,\tau}-M_{out}X + M_{out}QN) z^\tau \right\|_2^2 = 0.
\end{equation}
Thus, it holds that
\begin{equation*}
\begin{aligned}
\inf\limits_{Q \in \mathcal{RH}_\infty}\left\| W(Y + M Q)N \right\|_2^2=\left\| Z_2\right\|_2^2.
\end{aligned}
\end{equation*}

To obtain the expression of $Z_2$, let the impulse response of $W$ be $\{w(0), w(1), \cdots\}$, i.e.,
\begin{align}\label{W_z_transformation}
W=w(0)+w(1)z^{-1}+w(2)z^{-2}+\cdots.
\end{align}

From the state-space model of $M_{in}^{-1}$ in (\ref{M_in_inv_realization}), it holds that
\begin{align}\label{M_in_invs_N}
M_{in}^{-1}&=D_{in}^{-1}+D_{in}^{-1}C_{in}A_{in}^*B_{in}D_{in}^{-1}\\
&\hspace{1.2cm}+D_{in}^{-1}C_{in}{A_{in}^*}^2B_{in}D_{in}^{-1}z+D_{in}^{-1}C_{in}{A_{in}^*}^3B_{in}D_{in}^{-1}z^2+\cdots
\nonumber
\end{align}
and
\begin{align}\label{M_in_invs_P}
\hat{M}_{in,\tau}&=D_{in}^{-1}-D_{in}^{-1}C_{in}B_{in}D_{in}^{-1}z^{-1}\\
&\hspace{0.4cm}-D_{in}^{-1}C_{in}{A_{in}^{-*}}B_{in}D_{in}^{-1}z^{-2}-\cdots
-D_{in}^{-1}C_{in}{A_{in}^{-*}}^{\tau-2}B_{in}D_{in}^{-1}z^{-\tau+1}.\nonumber
\end{align}
Following (\ref{M_in_invs_N}) and (\ref{M_in_invs_P}), we obtain that
\begin{align}\label{W_2_1}
(M_{in}^{-1}-\hat{M}_{in,\tau})z^{\tau}
&=D_{in}^{-1}C_{in}{A_{in}^*}^{-\tau+2}B_{in}D_{in}^{-1}z+\cdots
+D_{in}^{-1}C_{in}B_{in}D_{in}^{-1}z^{\tau-1}\\
&\hspace{0cm}+D_{in}^{-1}C_{in}A_{in}^*B_{in}D_{in}^{-1}z^{\tau}
+D_{in}^{-1}C_{in}{A_{in}^*}^2B_{in}D_{in}^{-1}z^{\tau+1}\nonumber\\
&+D_{in}^{-1}C_{in}{A_{in}^*}^3B_{in}D_{in}^{-1}z^{\tau+2}+\cdots.\nonumber
\end{align}

Since $Z_2 \in \mathcal{RH}_2^\perp$, it holds that
\begin{align}
Z_2=\sum_{k=1}^{\infty}z_2(k)z^k
\end{align}
where
\begin{align}\label{z_k}
z_2(k)=\frac{1}{2\pi}\oint_{\partial \mathbb{D}}z^{-k-1}W(M_{in}^{-1}-\hat{M}_{in,\tau})z^{\tau}dz.
\end{align}
Applying (\ref{W_z_transformation}) and (\ref{W_2_1}) to (\ref{z_k}) yields
that
\begin{align}
z_2(k)&=\sum_{i=0}^{\infty}D_{in}^{-1}C_{in}{A_{in}^{-*}}^{\tau-1-k-i}B_{in}D_{in}^{-1}w(i)\\
&=D_{in}^{-1}C_{in}W(A_{in}^{-*}){A_{in}^{-*}}^{\tau-1-k}B_{in}D_{in}^{-1}.\nonumber
\end{align}
Consequently, it holds that
\begin{align}\label{Z_2_H2_Norm}
\|Z_2\|_2^2=&\sum_{k=1}^{\infty}\left\{D_{in}^{-1}C_{in}W(A_{in}^{-*}){A_{in}^{-*}}^{\tau-1-k}B_{in}D_{in}^{-1}
\right.\\
&\hspace{2cm}\left. \times D_{in}^{-*}B_{in}^*{A_{in}^{-1}}^{\tau-1-k}W(A_{in}^{-1})C_{in}^*D_{in}^{-*}\right\}.\nonumber
\end{align}

On the other hand, it follows from (\ref{Balanced_realization}) that
\[
A_{in}^*B_{in}D_{in}^{-1}=-C_{in}^*, \quad C_{in}^*C_{in}=I-A_{in}^*A_{in}.
\]
Thus, we have
\begin{align}\label{Balanced_realization_1}
A_{in}^*B_{in}D_{in}^{-1}D_{in}^{-*}B_{in}^*A_{in}=I-A_{in}^*A_{in}.
\end{align}
Substituting (\ref{Balanced_realization_1}) into (\ref{Z_2_H2_Norm})
\begin{align*}
\inf\limits_{Q \in \mathcal{RH}_\infty}\left\| W(Y + M Q)N \right\|_2^2
=&D_{in}^{-1}C_{in}W(A_{in}^{-*}){A_{in}^{-*}}^{\tau-1}A_{in}^{-(\tau-1)}W(A_{in}^{-1})C_{in}^*D_{in}^{-*}.
\end{align*}
The proof is completed by using Lemma \ref{Lma:Stabilizable_condition}.
\end{proof}

{
\begin{remark}\label{Remark:Qopt}
The controller which solves the minimization problem in Lemma 4.2 can be obtained from \eqref{Equ:Youla_param_controller} with $Q$ being such that \eqref{Equ:Optimal_Q} holds.
Thus, if the networked system is stabilizable, i.e., \eqref{Equ:Existence_condition} holds, then a stabilizing controller is straightforward.
\end{remark}
}

Now, several special cases of this theorem are discussed.

\begin{cor}\label{packet_drop}
{Suppose that the relative degree $\tau$ of the plant $P$ is one and the channel uncertainty is induced by random packet dropout with a given rate $p$. Then, the frequency response of variation $W$ of the channel is a constant.
The networked system is mean-square input-output stabilizable if and only if
\begin{align*}
\frac{1}{2}\log\left(1+\frac{1}{W^2} \right) > \sum_{i=1}^n\log|\lambda_i|
\end{align*}
or
\begin{align*}
p < \prod_{i=1}^n |\lambda_i|^{-2}.
\end{align*}
}
\end{cor}

\begin{proof}
{Since the channel uncertainty is induced by random packet drop only, it holds for the channel model shown in Fig. \ref{Fig:System_connected_channel} that $\bar{\tau}=1$ and $\alpha_1=0$. Subsequently, we have $H=\alpha_0\, (1-p)$ and $S_{\Omega}=\alpha_0^2\, p\, (1-p)$. This leads to ${\displaystyle W=\sqrt{\frac{p}{1-p}}}$.}

Notice that $W(A_{in}^{-*})=WI$, where $I$ is an identity matrix, since the {{frequency response of variation
$W$}} is a scalar constant. In this case, the inequality (\ref{Equ:Existence_condition}) is written as
\begin{align}\label{W_constant}
D_{in}^{-1}C_{in}C_{in}^*D_{in}^{-*}<\frac{1}{W^2}.
\end{align}
Following Lemma \ref{Balanced_realization_Lemma}, we have
\begin{align}\label{Cin_Din}
C_{in}C_{in}^*+D_{in}D_{in}^*=1.
\end{align}
Substituting (\ref{Cin_Din}) into (\ref{W_constant}) leads to
\begin{align*}
D_{in}^{-1}D_{in}^{-*}-1<\frac{1}{W^2}.
\end{align*}
Moreover, it follows from (\ref{Equ:Factorize_M}) that
\[
D_{in}=M_{in}(\infty)=\prod_{i=1}^n (-\lambda_i^{-*}).
\]
Consequently, according to Theorem \ref{Thm:Mean_squre_bounds}, the networked system is mean-square input-output stabilizable
if and only if it holds that
\begin{align*}
\prod_{i=1}^n |\lambda_i|^2<1+\frac{1}{W^2}.
\end{align*}
Thus, this corollary holds.
\end{proof}


\begin{cor}\label{Coro:One_unstable_pole}
Suppose that the plant $P$ has only one unstable pole $\lambda$ and is with the relative degree one, i.e., $\tau=1$.
The networked system \hlt{with a random channel induced delay and a given linear receiving strategy} is mean-square input-output stabilizable if and only if
\begin{align}\label{One_lambda}
\frac{1}{2}\log\left(1+\frac{1}{W^2(\lambda)} \right) > \log|\lambda|.
\end{align}
\end{cor}

\begin{proof}
Since $P$ has only one unstable pole $\lambda$, the inner of $M_{in}={\displaystyle
\frac{z-\lambda}{1-\lambda^*z}}$ is given by
\begin{align}\label{Bal_real_one_factor}
M_{in}=\sys{\displaystyle \frac{1}{\lambda^*}}{{\displaystyle \frac{\sqrt{\lambda
\lambda^*-1}}{\lambda^*}}}{{\displaystyle \frac{\sqrt{\lambda\lambda^*-1}}{\lambda^*}}}{-\displaystyle
\frac{1}{\lambda^*}}.
\end{align}
Noting the facts that $\tau=1$ and $\lambda$ is a real number, we have that
\begin{align*}
&D_{in}^{-1}C_{in}{A_{in}^{-*}}^{\tau-1}W(A_{in}^{-*}) W(A_{in}^{-1}) A_{in}^{-(\tau-1)} C_{in}^*D_{in}^{-*}
=(\lambda^2-1)W^2(\lambda).
\end{align*}
From Theorem \ref{Thm:Mean_squre_bounds}, the networked system is mean-square input-output stabilizable if and only if the inequality (\ref{One_lambda}) holds.
\end{proof}

\section{Numerical Examples}\label{Sec:Examples}

In this section, we illustrate the reason that weights should be assigned to the received signals and verify the
stabilizability criterion given in Theorem \ref{Thm:Mean_squre_bounds} by numerical examples.

\subsection{Weighting the received signals} 

Consider a discrete-time LTI plant
\begin{equation}\label{Equ:Example_plant}
P = \frac{z+0.9}{(z+1.2) (z-1.1)}
\end{equation}
connected with a one-step random delay channel. The PMF of the random delay is $p_0 ={\displaystyle  \frac{5}{11}}$ and $p_1=
{\displaystyle \frac{6}{11}}$.
Without assigning any weights to the received data, the mean channel would be
\begin{equation*}
H(z) = \frac{5}{11} + \frac{6}{11}z^{-1} = \frac{5z + 6}{11 z},
\end{equation*}
which has a non-minimum phase zero, $z_0 = -1.2$, coincided with one of the unstable poles of the plant.
Therefore, there occurs unstable pole-zero cancelation in the nominal closed-loop system $G$. Consequently, it
holds for any controller $K$ that
\begin{equation*}
\begin{aligned}
G(z) &= KP(1-HKP)^{-1}\\
&= \frac{{z K(z)\left( {z + 0.9} \right)}}{{\left( {z + 1.2} \right)\left[ {z\left( {z - 1.1} \right) -  0.45 K(z)\left(
{z + 0.9} \right)} \right]}}
\end{aligned}
\end{equation*}
where the unstable pole at $z = -1.2$ can not be changed by designing a proper controller. The closed-loop system is not
stabilizable.

To avoid the unstable pole-zero cancelation, we assign a set of weights, say $\alpha_0 = 0.8$ and $\alpha_1 =
0.2$, to the received data. Then the mean channel becomes $H(z) = {\displaystyle \frac{40z + 12}{110z}}$.
This prevents the cancelation between the zero of $H(z)$ and the unstable pole of the plant
\eqref{Equ:Example_plant} and makes it possible to stabilize the networked feedback system.

\subsection{Mean-square stabilizability index}

\hlt{

Consider a networked feedback system whose plant is a discrete-time LTI minimum phase system
\begin{equation}\label{Exam:P}
P = \frac{z - 0.2}{z^r(z-1.1)(z-1.2)},~~r \ge 0
\end{equation}
The relative degree of the plant is $\tau = r+1$.
In the networked system, the channel induced delay is characterized by $\bar{\tau} = 2$ with delay probabilities $p_0 = 0.6,~ p_1 = 0.3$
and packet loss probability $p_2 = 0.1$.
The weights to the received data are set as $\alpha_0 = 0.6,~\alpha_1  = 0.4,~\alpha_2  = 0$.
Under this setting, the mean channel $H(z)$ is minimum phase and the {{frequency response of variation}}, which only
depends on the channel and receiver, is given by
\begin{align*}
\hlt{ W(z)= \frac{0.3188 - 0.1355 z^{-1} + 0 z^{-2}}{0.36 + 0.12 z^{-1} + 0z^{-2}} = \frac{0.8856 (z - 0.425)}{z+0.3333}  }
\end{align*}

Let the coprime factorization $HP=NM^{-1}$ be
\begin{equation*}
N = \frac{0.12(3z+1)(z - 0.2)}{z^{r+1}(1-1.1z)(1-1.2z)},~M=\frac{(z-1.1)(z-1.2)}{(1-1.1z)(1-1.2z)}
\end{equation*}
It is verified that $M$ is inner, i.e.,
$$M_{in} = \frac{(z-1.1)(z-1.2)}{(1-1.1z)(1-1.2z)}.$$
The balance realization of $M_{in}$ is
\begin{equation*}
\begin{aligned}
A_{in} &= \begin{bmatrix}
    0.7500  & -0.1144\\
    0.1159  &  0.9924
         \end{bmatrix},
B_{in} = \begin{bmatrix}
   -0.6515\\
   -0.0408
         \end{bmatrix},\\
C_{in} &= \begin{bmatrix}
    0.6512 &  -0.0449
         \end{bmatrix},
D_{in} = 0.7576,
\end{aligned}
\end{equation*}
which obviously satisfies \eqref{Balanced_realization}.

According to Theorem \ref{Thm:Mean_squre_bounds}, the term on the left hand side of \eqref{Equ:Existence_condition} indicates the mean-square input-output stabilizability of a networked feedback system.
Once it is greater than one, no controller can stabilize the system in mean-square sense, i.e., the system is not mean-square input-output stabilizable. Here, we use this term as a {\em mean-square stabilizability index} for the system.
For the plant with $r=0$, namely $\tau=1$, this mean-square stabilizability index is given by
\begin{equation}\nonumber
D_{in}^{-1}C_{in}W(A_{in}^{-*}) W(A_{in}^{-1}) C_{in}^* D_{in}^{-*} = 0.1728,
\end{equation}
which implies that the networked system can be stabilized by some controller in the mean-square sense.
As shown by \eqref{Equ:Existence_condition}, the mean-square stabilizability index is also related to the relative degree $\tau$ of the plant $P$. Since all the eigenvalues of $A_{in}^{-1}$ are outside the unit disk, the mean-square stabilizability index in exponentially increases with respect to the relative degree.
In this example, the stabilizability index with respect to the relative degree $\tau$ is shown in Fig. \ref{Fig:Relative_degree}.
When the relative degree $\tau$ grows to $5$, the mean-square stabilizability index is greater than one and the
system is not mean-square stabilizable.

\begin{figure}[!htb]
  \centering
  \includegraphics[width=8cm]{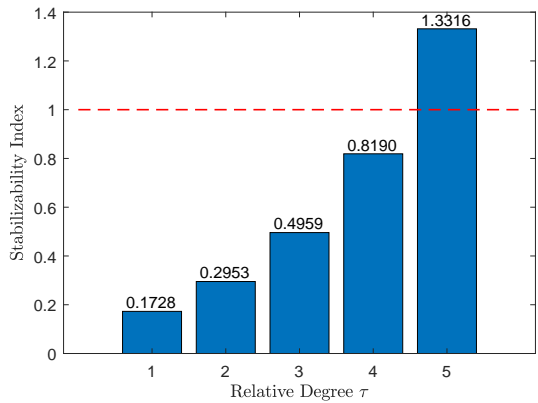}
  \caption{Relative degree vs. stabilizability index}\label{Fig:Relative_degree}
\end{figure}

To verify Theorem \ref{Thm:Mean_square_IO_stable}, we also take $r=0$ in the plant \eqref{Exam:P} as an example.
Since the mean-square input-output stability of the system means the boundedness of the average power $\|u\|_{\cal P}^2$ of the control signal $u$, it would be suitable to use the graph of $\|u\|_{\cal P}^2$ against $\left\|WT\right\|_2^2$ to visualize the stability or instability of the closed-loop system.
Noticing that the networked system is mean-square input-output stabilizable, from Remark \ref{Remark:Qopt}, a straightforward stabilizing controller of the networked system should be that in \eqref{Equ:Youla_param_controller} with $Q=Q_{opt}$ where $Q_{opt}$ is the solution to the equation \eqref{Equ:Optimal_Q}.
In this case, $\|WT\|_2^2$ achieves its minimum, i.e., the stabilizability index, $0.1728$.
Now let the controller $K $ be in $\mathcal{K}$ of \eqref{Equ:Youla_param_controller} with $Q = Q_{opt}+ \kappa \tilde{Q}$ where $\tilde{Q} \in \mathcal{RH}_{\infty}$ and $\kappa$ is a real nonnegative number.
It turns out that, for a given $\tilde{Q}$, by varying $\kappa$ from 0 to some sufficiently large number (dependent on $\tilde{Q}$), $\|WT\|_2^2$ would range from the stabilizability index to 1 such that the system is eventually unstable.
For seeing this, three different $\tilde{Q}$'s are taken into consideration, as shown in Fig. \ref{Fig:MS_simulation}.
By the Monte-Carlo method, the graphs of the theoretical and simulated average powers $\|u\|_{\cal P}^2$ against $\|WT\|_2^2$ associated with the $\tilde{Q}$'s are also illustrated in Fig. \ref{Fig:MS_simulation}, provided that the external input of the closed-loop system is a zero-mean Gaussian white noise with unit-variance.
All average powers of the control signals would tend to infinity as $\|WT\|_2^2$ approaches $1$.
This confirms Theorem \ref{Thm:Mean_square_IO_stable} and implies that improper controller design would destabilize the closed-loop system. 

\begin{figure}[!htb]
  \centering
  \includegraphics[width=8cm]{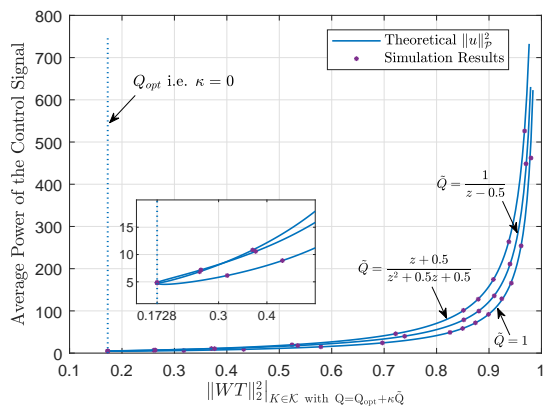}
  \caption{Average powers of the control signals vs. $\|WT\|_2^2$}\label{Fig:MS_simulation}
\end{figure}


}

\section{Conclusion}\label{Sec:Conclusion}
In this paper, we have studied the \msis and stabilizability for a discrete-time LTI networked
feedback system over an unreliable channel with random channel induced delays and packet dropout.
\hlt{Under a given linear receiving strategy}, the models of the unreliable channel and related channel uncertainty are presented in time domain and frequency domain,
respectively. In particular, {{frequency response of variation}} is introduced to describe the relative derivation of
the unreliable channel. Applying these models, the \msis {criterion} is obtained for the networked feedback system. This is a general version of the mean-square small gain theorem for discrete-time LTI systems
with i.i.d. stochastic multiplicative uncertainties.
After then, the mean-square stabilizability is studied for the networked feedback system when its plant is minimum
phase.
A necessary and sufficient condition is found for the mean-square stabilizability of the networked feedback system
via output feedback.
This result shows the inherent connection between the mean-square stabilizability, the plant's unstable poles and the frequency response of variation of the channel in the system.

{
It is straightforward to apply the proposed stability results to the so-called two-side networked control systems, i.e., systems with signal transmission via networks on both sensor and actuating channels. However, it can be shown that the stabilizability problem would become a decentralized control problem with information constraints, which precludes the convexity of the problem.
Thus, exploring the stabilizability criteria for this type of systems will be focus of future research.
It is also natural to consider the stability of the networked feedback system over the unreliable channel with deterministic constraints from the plant or/and the channel, which may turn out to be a mixed problem.
\hlt{Moreover, future work should include the stabilizability problems in a general setup of designable linear receiving strategy.}
}



\end{document}